\documentclass[12pt,a4paper]{article}
\usepackage{graphicx}
\usepackage[T1]{fontenc}
\usepackage[utf8]{inputenc}
\usepackage{textcomp}
\usepackage[sc,osf]{mathpazo}
\usepackage{a4wide}  
\usepackage{latexsym,amsthm,amsfonts,amsmath,mathrsfs,amssymb}
\usepackage{accents}
\usepackage[nosort]{cite}
\usepackage{booktabs} 
\usepackage[unicode,implicit]{hyperref}
\hypersetup{%
  pdftitle    = {The Komar charges of type~II supergravity and supersymmetric localization}
  pdfkeywords = {gravity, supergravity, symmetries, unbroken supersymmetry, conserved charges, Komar charge},
  pdfauthor   = {Gabriele Barbagallo and Tom\'as Ort\'{\i}n},
  plainpages  = true,
  colorlinks  = true,
  citecolor   = blue,
  urlcolor    = red,
  linkcolor   = black
}
\newcommand{\hepth}[1]{{\tt
\href{http://www.arXiv.org/abs/hep-th/#1}{hep-th/#1}}}
\newcommand{\grqc}[1]{{\tt
\href{http://www.arXiv.org/abs/gr-qc/#1}{gr-qc/#1}}}

\newcommand{\matharxiv}[1]{{\tt
\href{http://www.arXiv.org/abs/math/#1}{math/#1}}}
\newcommand{\arxiv}[1]{{\tt arXiv:\href{http://www.arXiv.org/abs/#1}{#1}}}
\newcommand{\doi}[1]{{\tt DOI:\href{http://dx.doi.org/#1}{#1}}}

\allowdisplaybreaks

\makeatletter
\@addtoreset{equation}{section}
\makeatother

\begin{document}

\begin{flushright}
\small
IFT-UAM/CSIC-26-131\\
\texttt{arXiv:2610.10045 [hep-th]}\\
October 8\textsuperscript{th}, 2026\\
\normalsize
\end{flushright}

\vspace{1cm}

\begin{center}

  {\Large {\bf The Komar charges of type~II supergravity\\[.5cm] and supersymmetric localization}}
 
\vspace{1cm}

\renewcommand{\thefootnote}{\alph{footnote}}

{\sl Gabriele Barbagallo}\footnote{Email: {\tt gabriele.barbagallo[at]estudiante.uam.es}}
{\sl and Tom\'as Ort\'{\i}n}\footnote{Email: {\tt Tomas.Ortin [at] csic.es}}\\[.5cm]

\setcounter{footnote}{0}
\renewcommand{\thefootnote}{\arabic{footnote}}
\vspace{.7cm}

${}^{1}${\it Instituto de F\'{\i}sica Te\'orica UAM/CSIC\\
  C/ Nicol\'as Cabrera, 13--15, C.U.~Cantoblanco, E-28049 Madrid, Spain}

\vspace{1.5cm}

{\bf Abstract}
\end{center}
\begin{quotation} {\small We compute the on-shell closed, generalized Komar
    charges associated to an arbitrary Killing vector of the two
    10-dimensional $\mathcal{N}=2$ supergravity theories which are the
    low-energy effective field theories of the type~IIA and IIB superstrings.
    We study different rewritings of this charge in terms of Page charges and
    forms with special properties that are used in the derivation of
    thermodynamical relations in black $p$-brane spacetimes. Then, we show
    that, when the Killing vector is the supersymmetric Killing vector built
    as a bilinear of a Killing spinor, the momentum maps of the different
    fields of the theory are given by other Killing spinor bilinears and that
    the differential equation satisfied by the supersymmetric Killing vector
    is, then, equivalent to the vanishing of the associated generalized Komar
    charge.}
\end{quotation}

\newpage
\pagestyle{plain}

\tableofcontents

\newpage

\section{Introduction}

It is difficult to overstate the role that conserved charges play in physics.
Our current understanding is that, in theories that include gravity, the
relevant conserved charges are associated to local symmetries and are
expressed as integrals of $(d-2)$-form charges over the boundary at spatial
infinity \cite{Regge:1974zd}. These are, in fact, the only charges which are
conserved when quantum gravity effects are taken into account since these are
expected to break all global symmetries \cite{Banks:1988yz,Banks:2010zn}.

In general, the $(d-2)$-form charges associated to gravitational charges
\cite{Abbott:1981ff} or charges of non-Abelian theories \cite{Abbott:1982jh}
do not satisfy a Gauss law,\footnote{See Ref.~\cite{Barnich:2001jy} a
  comprehensive approach.} but, in special circumstances, it is possible to
define $(d-2)$-form charges which are closed (conserved) in the bulk of
spacetime and, therefore, satisfy a Gauss law \cite{Barnich:2003xg}. The prime
example of this kind of charges is the Komar charge of pure General Relativity
(GR) \cite{Komar:1958wp}, further generalized in
Refs.~\cite{Bardeen:1973gs,Carter:1973rla,Magnon:1985sc,Bazanski:1990qd,Kastor:2008xb},
associated to a Killing vector, which is closed on-shell. Thus, they are
conserved only for solutions of the equations of motion which have a symmetry
but, in spite of this limitation, they have shown to be extremely useful to
derive generalizations of the Smarr formula \cite{Smarr:1972kt} in black-hole
spacetimes and no-hair theorems \cite{Ballesteros:2023iqb} or to study the
existence of singularity-free (solitons, boson stars) solutions
\cite{Ballesteros:2024prz,Barbagallo:2026hof}. Hence the interest in them.

In Refs.~\cite{Liberati:2015xcp,Ortin:2021ade,Ortin:2022uxa,Cerdeira:2025elp}
a systematic procedure to construct generalized Komar charges (conserved in
solutions admitting a Killing vector that generates a symmetry of all the
fields) has been devised. Using this procedure, the generalized Komar charge
of different theories in different dimensions
\cite{Mitsios:2021zrn,Ortin:2022uxa,Meessen:2022hcg,Ballesteros:2023muf,Barbagallo:2025qdy,Barbagallo:2026hof}
and with the inclusion of fermionic fields and supersymmetry
\cite{Bandos:2023zbs,Bandos:2024rit} have been derived.

It turns out that some of the objects that naturally occur in the equivariant
localization approach to the calculation of the Euclidean action for
supersymmetric solutions
\cite{BenettiGenolini:2023kxp,Couzens:2026xmi,BenettiGenolini:2026kjy} seem to
be related to the are related to the generalized Komar charge, although this
relation has not yet been made precise.

On the other hand, the generalized Komar charge associated to the
supersymmetric Killing vector of the supersymmetric solutions of certain
supergravity theories has been shown in Ref.~\cite{Ortin:2026fzm} to vanish
identically. This is a remarkable property that allows one to prove the
saturation of BPS bounds, for instance, and that, should the relation with the
equivariant localization approach be confirmed, would have an important impact
in the latter. However, as yet, we do not know any reason why this property
should hold in all supergravity theories and, for the moment, we have to
content ourselves with the derivation and study of the generalized Komar
charges of more supergravity theories.

In Ref.~\cite{Couzens:2026xmi} the supersymmetric equivariant localization
approach has been applied to the $\mathcal{N}=2,d=10$ supergravity theories
which are the low-energy effective field theories of the type~IIA and IIB
superstrings. It is very natural to try to extend the results of
Ref.~\cite{Ortin:2026fzm} and in this work we are going to derive the
generalized Komar charges of these theories and investigate whether they
vanish for the supersymmetric Killing vector of supersymmetric solutions. We
will show that, indeed, this is the case when one identifies certain spinor
bilinears with the \textit{momentum maps}\footnote{These objects are
  differential forms that express the invariance of the fields under the
  transformations generated by a vector field that will be defined in the main
  text.} of the different fields. The proof of this result requires the
derivation of one differential equation for the supersymmetric Killing vector
that does not seem to be directly available in the literature \cite{Tomasiello:2011eb}.

Although, as we are going to see, we could have obtained the same results
using the democratic formulation of the $\mathcal{N}=2,d=10$ theories proposed
in Ref.~\cite{Bergshoeff:2001pv}, we have chosen to study study first the $2A$
and $2B$ theories separately as important warm-up exercises providing solid
results against which to compare those obtained with the democratic
formulation. Thus, in Section~\ref{sec-n2ad10sugra} we derive the generalized
Komar charge of the $2A$ theory, in Section~\ref{sec-n2ad10msugra} that of
Romans' massive $2A$ theory and in Section~\ref{sec-n2bd10sugra} that of the
$2B$ theory. The generalized Komar charge of the democratic theory is derived
in Section~\ref{sec-democratic} and it is shown how, when restricted to the
fields of the (massive) $2A$ and $2B$ theories one recovers the previous
results. Finally, in Section~\ref{sec-unbroken} we show that, for
supersymmetric solutions, the generalized Komar charge of the supersymmetric
Killing vector vanishes identically. Section~\ref{sec-discussion} contains our
conclusions and a series of questions concerning the meaning and generality of
our results we would like to investigate in future work.

Throughout this paper we use the conventions of
Ref.~\cite{Ortin:2015hya} for the (mostly minus) metric, vielbein,
spin connections, curvatures, differential forms and 10-dimensional
gamma matrices and spinors with the minor addition of the notation
$\imath_{a}\omega$ to denote the interior product of the basis vectors
$e_{a}=e_{a}{}^{\mu}\partial_{\mu}$ with the differential form
$\omega$.

\section{The generalized Komar charge of $\mathcal{N}=2A,d=10$ supergravity}
\label{sec-n2ad10sugra}

The bosonic sector of the $\mathcal{N}=2A,d=10$ supergravity theory
\cite{Giani:1984wc,Campbell:1984zc} consists of 3 Neveu--Schwarz-Neveu--Schwarz
(NSNS) fields: the Zehnbein, $e^{a}$, the dilaton, $\phi$ and the Kalb--Ramond
(KR) 2-form potential, $B$, and 2 Ramond-Ramond (RR) fields: the 1- and 3-form
potentials $C^{(1)}$ and $C^{(3)}$.

In the string frame and in differential-form language, the action for these
fields can be brought to the form

\begin{subequations}
  \label{eq:n2ad10action}
  \begin{align}
    S 
    & =
    \frac{g_{s\, A}^{2}}{16\pi G_{N\, A}^{(10)}}
      \int_{\mathcal{M}_{10}} \left\{
      \mathbf{L}_{NSNS} +\mathbf{L}_{RR}
      \right\}\,,
    \\
    & \nonumber \\
    \label{eq:LNSNS}
    \mathbf{L}_{NSNS}
    & \equiv
      e^{-2\phi}
    \left[ -\star (e^{a}\wedge e^{b}) \wedge R_{ab}
    -4d\phi\wedge \star d\phi
    +\tfrac{1}{2}H\wedge \star H
      \right]\,,
    \\
    & \nonumber \\
        \label{eq:LRR2A}
    \mathbf{L}_{RR}
    & =
      \tfrac{1}{2}G^{(2)}\wedge \star G^{(2)}
      +\tfrac{1}{2}G^{(4)}\wedge \star G^{(4)}
      \nonumber    \\
    & \nonumber \\
    & \hspace{.5cm}
      -\tfrac{1}{2}G^{(4)}\wedge G^{(4)}\wedge B
      +\tfrac{1}{2}G^{(4)}\wedge G^{(2)} \wedge B^{2}
      -\tfrac{1}{3!}G^{(2)}\wedge G^{(2)} \wedge B^{3}\,,
  \end{align}
\end{subequations}

\noindent
where $H,G^{(2)}$ and $G^{(4)}$ are, respectively, the gauge-invariant field
strengths of $B,C^{(1)}$ and $C^{(3)}$, defined as

\begin{equation}
  H\equiv dB\,,
  \hspace{1cm}
  G^{(2)} \equiv dC^{(1)}\,,
  \hspace{1cm}
  G^{(4)} \equiv dC^{(3)}-H\wedge C^{(1)}\,.
\end{equation}

The gauge transformations that leave them invariant are

\begin{equation}
  \label{eq:gaugetransformations2A}
\delta_{\Lambda} B = d\Lambda^{(1)}_{B}\,,   
  \hspace{1cm}
\delta_{\Lambda} C^{(1)} = d\Lambda^{(0)}\,,   
  \hspace{1cm}
\delta_{\Lambda} C^{(3)} = d\Lambda^{(2)}-H\Lambda^{(0)} \,.   
\end{equation}

We are using the standard notation $B^{2}\equiv B\wedge B$,
$B^{3}\equiv B\wedge B\wedge B$ etc. The Chern-Simons term of this action (the
second line of Eq.~(\ref{eq:LRR2A})) is equivalent, up to total derivatives, to
the more common expression

\begin{equation}
-\tfrac{1}{2}dC^{(3)}\wedge dC^{(3)}\wedge B\,,  
\end{equation}

\noindent
but, unlike it, it is manifestly invariant under the RR gauge transformations
with parameters $\Lambda^{(0)}$ and $\Lambda^{(2)}$ and, thus, we only have to
worry about the KR gauge transformations with gauge parameter
$\Lambda^{(1)}_{B}$. These leave it invariant up to a total derivative:

\begin{equation}
  \label{eq:deltaLambdaS2A}
  \begin{aligned}
    \delta_{\Lambda}S
    & =
      -\int_{\mathcal{M}_{10}} d\mathbf{Y}^{(7)}\wedge \delta_{\Lambda}B\,,
    \\
    & \\
    \mathbf{Y}^{(7)}
    & \equiv
      \tfrac{1}{2}d(C^{(3)}-B\wedge C^{(1)}) \wedge (C^{(3)}-B\wedge C^{(1)})\,.
    \\
    & \\
    d\mathbf{Y}^{(7)}
    & \equiv
      \tfrac{1}{2}\left(G^{(4)}-B\wedge G^{(2)} \right)
      \wedge \left(G^{(4)}-B\wedge G^{(2)} \right)\,.
  \end{aligned}
\end{equation}

The normalization of the action in the string frame contains the $IIA$ string
coupling constant $g_{s\, A}$ and the $IIA$ 10-dimensional Newton constant
$G_{N\, A}^{(10)}$ but, in what follows, we will ignore this factor in order
to get simpler expressions.

Under an arbitrary infinitesimal variation of the fields, the action changes
as

\begin{equation}
  \label{eq:generalvariation2a}
  \delta S
  =
  \int_{\mathcal{M}_{10}} \left\{
    \mathbf{E}_{a}\wedge \delta e^{a}
    +\mathbf{E}_{\phi}\delta \phi
    +\mathbf{E}_{B}\wedge \delta B
    +\mathbf{E}_{(1)}\wedge \delta C^{(1)}
    +\mathbf{E}_{(3)}\wedge \delta C^{(3)}
    +d\mathbf{\Theta}(\varphi,\delta \varphi)
  \right\}\,,  
\end{equation}

\noindent
where, by definition,

\begin{subequations}
  \label{eq:eom2A}
  \begin{align}
    \mathbf{E}_{a}
    & =
      e^{-2\phi}\left[\imath_{a}\star (e^{b}\wedge e^{c})\wedge R_{bc}
      -4(\imath_{a}d\phi \star d\phi
      +d\phi\wedge \imath_{a} \star d\phi) \right.
      \nonumber \\
    & \nonumber \\
    & \hspace{.5cm}
      \left.
      +\tfrac{1}{2}(\imath_{a}H \wedge\star H
      +H\wedge \imath_{a} \star H)
      +2e^{2\phi}\mathcal{D}\left(\imath_{a}\star de^{-2\phi}\right)\right]
    \nonumber \\
    & \nonumber \\
    & \hspace{.5cm}
            +\tfrac{1}{2}(\imath_{a}G^{(2)} \wedge\star G^{(2)}
      -G^{(2)}\wedge \imath_{a} \star G^{(2)})
    \nonumber \\
    & \nonumber \\
    & \hspace{.5cm}
            +\tfrac{1}{2}(\imath_{a}G^{(4)} \wedge\star G^{(4)}
      -G^{(4)}\wedge \imath_{a} \star G^{(4)})\,,
      \\
    & \nonumber \\
    \mathbf{E}_{\phi}
    & =
      -4d\star de^{-2\phi} -2\mathbf{L}_{NSNS}\,,
    \\
    & \nonumber \\
    \mathbf{E}_{B}
    & =
      -d(e^{-2\phi}\star H) +\star G^{(4)} \wedge G^{(2)}
      -\tfrac{1}{2} G^{(4)} \wedge G^{(4)} -\mathbf{E}_{(3)}\wedge C^{(1)}\,,
    \\
    & \nonumber \\
    \mathbf{E}_{(1)}
    & =
      -d\star G^{(2)} -H \wedge \star G^{(4)}\,,
    \\
    & \nonumber \\
    \mathbf{E}_{(3)}
    & =
      -d\star G^{(4)} +H \wedge G^{(4)}\,,
  \end{align}
\end{subequations}

\noindent
are, respectively, the equations of motion of the Vielbein (the Einstein
equations), dilaton, KR and RR fields, and ($\varphi$ stands for all the
fields of the theory)

\begin{equation}
  \label{eq:Theta2A}
  \begin{aligned}
    \mathbf{\Theta}(\varphi,\delta \varphi)
    & =
    -e^{-2\phi}\star (e^{a}\wedge e^{b})\wedge \delta \omega_{ab}
    +4\star d e^{-2\phi} \delta\phi
      -2\imath_{a}\star d e^{-2\phi} \wedge \delta e^{a}
    \\
    & \\
    & \hspace{.5cm}
      +\mathbf{Z}^{(7)}\wedge \delta B
      +\mathbf{X}^{(8)}\wedge \delta C^{(1)}
      +\mathbf{X}^{(6)}\wedge \delta C^{(3)}\,,
  \end{aligned}
\end{equation}

\noindent
is the symplectic prepotential \cite{Lee:1990nz}.

We have defined

\begin{subequations}
  \begin{align}
    \mathbf{Z}^{(7)}
    & \equiv
      \frac{\delta S}{\delta H}
      =
      e^{-2\phi}\star H -\mathbf{X}^{(6)}\wedge C^{(1)}\,,
    \\
    & \nonumber \\
    \mathbf{X}^{(8)}
    & \equiv
      \frac{\delta S}{\delta G^{(2)}}
      =
      \star G^{(2)} +\tfrac{1}{2}G^{(4)}\wedge B^{2}
      -\tfrac{1}{3}G^{(2)}\wedge B^{3}\,, 
    \\
    & \nonumber \\
    \mathbf{X}^{(6)}
    & \equiv
      \frac{\delta S}{\delta G^{(4)}}
      =
      \star G^{(4)} -G^{(4)}\wedge B +\tfrac{1}{2} G^{(2)}\wedge B^{2}\,,
  \end{align}
\end{subequations}

\noindent
which satisfy the relations

\begin{subequations}
  \label{eq:XYZrelations2A}
  \begin{align}
    \label{eq:ZYrelation}
    d\left(\mathbf{Z}^{(7)} +\mathbf{Y}^{(7)}\right)
    & =
      -\mathbf{E}_{B}\,,
    \\
    & \nonumber \\
    \label{eq:X6relation}
    d\mathbf{X}^{(6)}
    & =
      -\mathbf{E}_{(3)}\,,
    \\
    & \nonumber \\
    \label{eq:X8relation}
    d\mathbf{X}^{(8)}+H\wedge \mathbf{X}^{(6)}
    & =
      -\mathbf{E}_{(1)}\,,
  \end{align}
\end{subequations}

\noindent
which simplifies the subsequent calculations.

\subsection{Symmetries and conserved charges}
\label{sec-n2ad10sugrasymmetriesandcharges}

\subsubsection{Local Lorentz transformations}
\label{sec-n2ad10LL}

The exact invariance of the action under local Lorentz transformations acting
on the Zehnbeins and spin connection as

\begin{equation}
  \delta_{\sigma}e^{a}
  =
  \sigma^{a}{}_{b}e^{b}\,,
  \hspace{1cm}
  \delta_{\sigma}\omega^{ab}
  =
  \mathcal{D}\sigma^{ab}\,,
\end{equation}

\noindent
has the symmetry of the Einstein equations as Noether identity and leads to the
off-shell conserved Noether current

\begin{equation}
  \begin{aligned}
    \mathbf{J}[\sigma]
    & =
      \mathbf{\Theta}(\varphi,\delta_{\sigma}\varphi)
    \\
    & \\
    & =
    -e^{-2\phi}\star (e^{a}\wedge e^{b})\wedge \mathcal{D} \sigma_{ab}
      -2\imath_{a}\star d e^{-2\phi} \wedge \sigma^{a}{}_{b} e^{b}
    \\
    & \\
    & =
      d\left\{ -e^{-2\phi}\star (e^{a}\wedge e^{b})\sigma_{ab} \right\}
    \\
    & \\
    & \equiv d \mathbf{Q}[\sigma]\,,
  \end{aligned}
\end{equation}

\noindent
where $\mathbf{Q}[\sigma]$ is the Noether charge associated to this symmetry.

Following the usual arguments, this charge should only be conserved for
Lorentz \textit{Killing} or \textit{reducibility} parameters
\cite{Barnich:2001jy} that leave invariant the Zehnbein and, therefore, leave
also invariant the connection.  This is impossible, though and such Killing
parameters do not exist. This seems to suggest that there are no conserved
Lorentz charges.

In the Einstein frame, though, if the Lorentz parameters are covariantly
constant and leave invariant the connection and not the Zehnbein, which is
possible, the Lorentz charge is conserved. We expect the same conserved
charges in the string and Einstein frames, and, indeed, it is not difficult to
see that the charge is conserved when the Lorentz parameter satisfies the
equation

\begin{equation}
  \mathcal{D}(e^{-2\phi}\sigma^{ab})
  =
  0\,.
\end{equation}

The Bekenstein--Hawking entropy of stationary black holes can be regarded as
the integral of the Lorentz charge associated to the Lorentz momentum map of
the Killing vector $k$ that generates the Killing horizon
$P_{k\, ab}=\mathcal{D}_{a}k_{b}$, rather than the integral of the
Noether--Wald charge as initially suggested in
Refs.~\cite{Wald:1993nt,Iyer:1994ys}. The reason is that, when there are
matter fields present, the latter includes other terms responsible for the
work terms in the first law of black hole mechanics.

\subsubsection{NSNS and RR gauge transformations}
\label{sec-n2ad10gauge}

It is convenient to study all these charges simultaneously
\cite{Barnich:2001jy,Barnich:2003xg}. Particularizing the generic variation of
the action Eq.~(\ref{eq:generalvariation2a}) to the gauge transformations
Eq.~(\ref{eq:gaugetransformations2A})

\begin{equation}
  \begin{aligned}
    \delta_{\Lambda} S
    & =
    \int_{\mathcal{M}_{10}} \left\{
    -d\mathbf{E}_{B}\wedge \Lambda^{(1)}_{B}
    +\left(d\mathbf{E}_{(1)}-\mathbf{E}_{(3)}\wedge H\right)\Lambda^{(0)}
      +d\mathbf{E}_{(3)}\wedge \Lambda^{(2)}
      \right.
    \\
    & \\
    & \hspace{.5cm}
      \left.
      +d\left[\mathbf{\Theta}(\varphi,\delta_{\Lambda} \varphi)
      +\mathbf{E}_{B}\wedge \Lambda^{(1)}_{B}
      -\mathbf{E}_{(1)}\wedge \Lambda^{(0)}
      -\mathbf{E}_{(3)}\wedge \Lambda^{(2)}\right]
      \right\}\,.
  \end{aligned}
\end{equation}

The terms in the first line vanish by virtue of Noether identities which are
easily checked using Eqs.~(\ref{eq:XYZrelations2A}).

On the other hand, as we have mentioned before, the action is only invariant
up to a total derivative under these transformations. Is is convenient to
write the total derivative in Eq.~(\ref{eq:deltaLambdaS2A}) in the form  

\begin{equation}
    \delta_{\Lambda}S
    =
      -\int_{\mathcal{M}_{10}}d\left[ \mathbf{Y}^{(7)}\wedge d\Lambda^{(1)}_{B}\right]\,,  
\end{equation}

\noindent
because the boundary term $\mathbf{Y}^{(7)}\wedge d\Lambda^{(1)}_{B}$ vanishes
identically when the gauge transformation generated by $\Lambda^{(1)}_{B}$
leaves invariant $B$, $d\Lambda^{(1)}_{B}=0$.

Comparing the two expression for $\delta_{\Lambda}S$ that we have obtained we
arrive at

\begin{equation}
  \begin{aligned}
    d\mathbf{J}[\Lambda]
    & =
      0\,,
    \\
    & \\
    \mathbf{J}[\Lambda]
    & =
    \mathbf{\Theta}(\varphi,\delta_{\Lambda} \varphi)
    +\mathbf{E}_{B}\wedge \Lambda^{(1)}_{B}
    -\mathbf{E}_{(1)}\Lambda^{(0)}
    -\mathbf{E}_{(3)}\wedge \Lambda^{(2)}
      +\mathbf{Y}^{(7)}\wedge d\Lambda^{(1)}_{B}\,.
  \end{aligned}
\end{equation}

Using the relations Eqs.~(\ref{eq:XYZrelations2A}) again, it is not difficult
to express $\mathbf{J}[\Lambda]$ as a total derivative

\begin{equation}
  \label{eq:QLambda2a}
  \begin{aligned}
    \mathbf{J}[\Lambda]
    & =
    d\mathbf{Q}[\Lambda]\,,
    \\
    & \\
    \mathbf{Q}[\Lambda]
    & =
 -(\mathbf{Z}^{(7)}+\mathbf{Y}^{(7)})\wedge  \Lambda^{(1)}_{B}
      +\mathbf{X}^{(8)}\Lambda^{(0)}
      +\mathbf{X}^{(6)}\wedge \Lambda^{(2)}
    \\
    & \\
    & =
 -(\mathbf{Z}^{(7)}+\mathbf{Y}^{(7)})\wedge  \Lambda^{(1)}_{B}
      +\left(\mathbf{X}^{(8)}+\mathbf{X}^{(6)}\wedge B\right)\Lambda^{(0)}
      +\mathbf{X}^{(6)}\wedge \left(\Lambda^{(2)} -B\Lambda^{(0)} \right)\,.
  \end{aligned}
\end{equation}

In the second line we have expressed $\mathbf{Q}[\Lambda]$ in terms of the
three independent combinations of parameters that leave the fields invariant,
$\Lambda^{(1)}_{B},\tilde{\Lambda}^{(0)}=\Lambda^{(0)},\tilde{\Lambda}^{(2)}=
\Lambda^{(2)} -B\Lambda^{(0)}$, when they are closed. The coefficients of
these parameters in $\mathbf{Q}[\Lambda]$ are closed on-shell
Eqs.~(\ref{eq:XYZrelations2A}), which ensures the on-shell conservation of
$\mathbf{Q}[\Lambda]$. Thus, we get three different charges

\begin{subequations}
  \begin{align}
    \mathbf{Q}[\Lambda^{(1)}_{B}]
    & =
      \Lambda^{(1)}_{B}\wedge (\mathbf{Z}^{(7)}+\mathbf{Y}^{(7)})\,,
    \\
    & \nonumber \\
    \mathbf{Q}[\tilde{\Lambda}^{(0)}]
    & =
\tilde{\Lambda}^{(0)}\left(\mathbf{X}^{(8)}+\mathbf{X}^{(6)}\wedge B\right)\,,    
    \\
    & \nonumber \\
    \mathbf{Q}[\tilde{\Lambda}^{(2)}]
    & =
      \tilde{\Lambda}^{(2)}\wedge \mathbf{X}^{(6)}\,,
  \end{align}
\end{subequations}

\noindent
which are non-trivially on-shell closed (conserved) when the gauge parameters
$\tilde{\Lambda}^{(0)}$, $\tilde{\Lambda}^{(2)}$ and $\Lambda^{(1)}_{B}$ are
closed but not exact.

On the other hand, those coefficients can be written in a systematic form if
we introduce the dual (``magnetic'') RR potentials $C^{(5)}$ and $C^{(7)}$
with field strengths\footnote{These definitions are not arbitrary. They follow
  from on-shell dualization.}

\begin{equation}
  G^{(6)} \equiv dC^{(5)} -H\wedge C^{(3)}\,,
  \hspace{1cm}
  G^{(8)} \equiv dC^{(7)} -H\wedge C^{(5)}\,,
\end{equation}

\noindent
and related to the fundamental (``electric'') ones by 

\begin{equation}
\label{eq:dualityconstraints}
G^{(10-n)} = (-1)^{[n/2]} \star G^{(n)}\,,
\end{equation}

\noindent
which will also be valid in the massive $2A$ and $2B$ theories. The Bianchi
identities of the dual potentials become the equations of motion of the
fundamental ones upon use of these relations and, therefore, we can always use
them on-shell.

Using electric and magnetic RR field strengths, then,

\begin{subequations}
  \begin{align}
    \mathbf{Z}^{(7)}+\mathbf{Y}^{(7)}
    & =
      e^{-2\phi}\star H
      -\tfrac{1}{2} \star G^{(8)}\wedge C^{(5)}
      -\tfrac{1}{2}\star G^{(6)}\wedge C^{(3)}
      -\tfrac{1}{2}\star G^{(4)}\wedge C^{(1)}   
      \nonumber \\
    & \nonumber \\
    & \hspace{.5cm}
      +d\left(  -\tfrac{1}{2}B\wedge C^{(1)}\wedge C^{(3)}
      -\tfrac{1}{2}C^{(5)}\wedge C^{(1)}\right)
    \\
    & \nonumber \\
    \mathbf{X}^{(6)}
    & =
      G^{(6)} -G^{(4)}\wedge B +\tfrac{1}{2} G^{(2)}\wedge B^{2}\,,
    \\
    & \nonumber \\
    \mathbf{X}^{(8)}+\mathbf{X}^{(6)}\wedge B
    & =
          -G^{(8)} +G^{(6)}\wedge B -\tfrac{1}{2}G^{(4)}\wedge B^{2}
      +\tfrac{1}{3!}G^{(2)}\wedge B^{3}\,.
   \end{align}
\end{subequations}

The RR combinations have a common structure that we can conveniently describe
using the polyforms\footnote{We also introduce the notation
  $\left. X\right|_{n}$ indicate the rank-$n$ term of the polyform $X$.}

\begin{equation}
  C \equiv \sum_{n=1,3,5,7}C^{(n)}\,,
\end{equation}

\noindent
and\footnote{In most of the literature $d_{H^{-}}$ is just denoted by
  $d_{H}$. However, we are going to use 
  \begin{equation}
    d_{H^{+}} \equiv d+H\wedge\,,
  \end{equation}
  as well. It occurs in the equations of motion that can be written in the form
  \begin{equation}
   d_{H^{+}}\star G =0\,. 
 \end{equation}
 Both derivatives are nilpotent when acting on polyforms
 \begin{equation}
 d_{H^{\pm}}^{2}=0\,.  
 \end{equation}
}

\begin{equation}
  G \equiv \sum_{n=2,4,6,8}G^{(n)}=\left(d-H\wedge\right) C \equiv d_{H^{-}}C\,,
\end{equation}

\noindent
through the on-shell exact but not gauge-invariant combinations

\begin{equation}
  \label{eq:Pagechargesdef}
\tilde{G} \equiv G \wedge e^{-B} = d\left(C\wedge  e^{-B}\right) \equiv dS\,,
\end{equation}

\noindent
which are nothing but the so-called \textit{Page charges} with associated Page
potentials

\begin{equation}
  S\equiv C\wedge  e^{-B}\,.
\end{equation}

\noindent
These are closed on-shell but not gauge invariant: defining the polyforms of
RR gauge parameters

\begin{equation}
  \Lambda
  \equiv
  \sum_{n=0,2,4,6}\Lambda^{(n)}\,,
  \hspace{1cm}
  \tilde{\Lambda}
  \equiv
  \Lambda\wedge e^{-B}\,,
\end{equation}

\noindent
in terms of which

\begin{equation}
\delta_{\Lambda} C = d_{H^{-}}\Lambda\,,  
\end{equation}

\noindent
we find

\begin{subequations}
  \label{eq:Pagechargesgaugetransformations}
  \begin{align}
    \delta_{\Lambda} S
    & =
      d\left[ (\Lambda -\Lambda_{B}^{(1)}\wedge C)\wedge e^{-B}\right]
      -\Lambda_{B}^{(1)}\wedge \tilde{G}\,,
    \\
    & \nonumber \\
    \delta_{\Lambda} \tilde{G}
    & =
      -d\left(\Lambda_{B}^{(1)}\wedge \tilde{G}\right)\,.
  \end{align}
\end{subequations}

Summarizing: the electric RR charges that we have found plus the magnetic ones
which we have not yet introduced can be systematically constructed (up to
global signs) in the form

\begin{equation}
  \label{eq:RRcharges2A}
  \mathbf{Q}[\tilde{\Lambda}^{(n)}]
  \equiv
  \tilde{\Lambda}^{(n)}\wedge \tilde{G}^{(8-n)}\,,
  \hspace{1cm}
  d\tilde{\Lambda}^{(n)}=0\,.
\end{equation}

Since the Page charges are gauge-invariant up to total derivatives, these charges
are also invariant up to total derivatives which are irrelevant when
integrated over boundaries.

As for the KR charge, we can define the dual, ``magnetic'' gauge-invariant
field strength $H^{(7)}$, the ``electric'' KR Page charge $\tilde{H}$ (equal
to $\mathbf{Z}^{(7)}+\mathbf{Y}^{(7)}$ up to total derivatives) and its
associated KR 6-form $B^{(6)}$ as

\begin{subequations}
  \begin{align}
    \label{eq:tildeHdef}
    \tilde{H} & \equiv
                e^{-2\phi}\star H -\tfrac{1}{2} \star G \wedge C \equiv dB^{(6)}\,,
    \\
    & \nonumber \\
    H^{(7)} & \equiv e^{-2\phi}\star H
              =
              dB^{(6)}+\tfrac{1}{2} \star G \wedge C\,.
  \end{align}
\end{subequations}

The dual KR 6-form transforms as


\begin{equation}
  \delta_{\Lambda} B^{(6)}
  =
  d\Lambda^{(5)}_{B}-\tfrac{1}{2} \star G \wedge \Lambda\,,  
\end{equation}

\noindent
and the charge associated to $\Lambda_{B}^{(1)}$ can be written in the form

\begin{equation}
  \label{eq:KRcharge2A}
  \mathbf{Q}[\Lambda^{(1)}_{B}]
  =
  \Lambda^{(1)}_{B}\wedge \tilde{H}\,,
  \hspace{1cm}
  d\Lambda^{(1)}_{B}=0\,.
\end{equation}

We can also introduce a ``magnetic'' KR charge 

\begin{equation}
  \mathbf{Q}[\Lambda^{(5)}_{B}]
  =
  \Lambda^{(5)}_{B}\wedge H \,,
  \hspace{1cm}
  d\Lambda^{(5)}_{B}=0\,,
\end{equation}

\noindent
although we are not going to need it in what follows.

\subsubsection{General coordinate transformations}
\label{sec-n2ad10gct}

As argued in Refs.~\cite{Elgood:2020nls,Elgood:2020mdx,Elgood:2020svt},
general coordinate transformations (GCTs) of fields with gauge freedoms are
not just given by the ordinary Lie derivative of tensor fields. It has to be
supplemented by an \textit{induced} or \textit{compensating} gauge
transformation $\delta_{\Lambda_{\xi}}$ whose gauge parameter $\Lambda_{\xi}$
depends on the vector field $\xi$ that generates the GCT. The generic form of
$\delta_{\xi}$ that we must consider is, then,

\begin{equation}
\delta_{\xi} = -\mathcal{L}_{\xi} +\delta_{\Lambda_{\xi}}\,.  
\end{equation}

If the field transforms under several local symmetries, there should be one
compensating gauge transformation of each of them.

When the transformation generated by a vector field $k$ leaves invariant the
field configuration\footnote{If one of the fields is the metric, then, in
  particular, that field is a Killing vector. We will call it Killing vector
  whenever it generates a GCT that leaves invariant the whole field
  configuration, whether or not the metric is part of it.} then $\delta_{k}$
defined above annihilates all the fields in a gauge-invariant way, eliminating
the need to fix a particular gauge that arises when one uses the standard Lie
derivative alone.\footnote{These ideas first arised in the construction of the
  so-called \textit{spinorial Lie derivatives} (or Lie-Lorentz derivatives
  acting on spinor fields) by Lichnerowicz and Kosmann
  \cite{kn:Lich,kn:Kos,kn:Kos2} (see also
  Refs.~\cite{Hurley:cf,Figueroa-O'Farrill:1999va}). The definition was
  extended to arbitrary Lorentz tensor-spinor fields in
  Ref.~\cite{Ortin:2002qb} and in Refs.~\cite{Godina:2003tc,Fatibene2011} a
  more rigorous mathematical construction was provided. In the context of
  Yang--Mills theories (which includes the Lorentz group as a particular case)
  general mathematical constructions have been given in
  Refs.~\cite{Prabhu:2015vua,Bunk:2023ojb}. A first application in the
  framework of black-hole thermodynamics was made in
  Ref.~\cite{Jacobson:2015uqa}. The fields and symmetries we are considering
  here (some of them, previously considered in
  Refs.~\cite{Elgood:2020mdx,Elgood:2020svt}) are not Yang--Mills fields and
  there is no rigorous fiber-bundle-type mathematical structure backing the
  somewhat \textit{ad-hoc} construction of $\delta_{\xi}$ that we are going to
  make, but its consistency is clear.}

As explained in Refs.~\cite{Elgood:2020nls,Elgood:2020mdx,Elgood:2020svt},
GCTs acting on Lorentz tensors or spinors induce a local Lorentz
transformation with parameter

\begin{equation}
  \label{eq:sigmaabk}
  \sigma_{\xi}{}^{ab}
  =
  \imath_{\xi}\omega^{ab} - P_{\xi}{}^{ab}\,,
\end{equation}

\noindent
where, for Killing vectors $k$, $P_{\xi}{}^{ab}$ is given by the
\textit{Killing bivector}

\begin{equation}
  \label{eq:Killingbivector}
P_{k}{}^{ab}=  \nabla^{a}k^{b}\,.
\end{equation}

This object satisfies the \textit{Lorentz momentum map equation} \cite{Elgood:2020nls},

\begin{equation}
  \label{eq:Lorentzmomentummapeq}
\imath_{k}R^{a}+\mathcal{D}  P_{k}{}^{ab}=0\,,
\end{equation}

\noindent
since it becomes equivalent to the integrability condition of the Killing
vector equation.

Using the above gauge transformation, we get

\begin{subequations}
  \begin{align}
    \label{eq:deltaxiea}
    \delta_{\xi}e^{a}
    & =
      -\left(\mathcal{D}\xi^{a}-P_{\xi}{}^{a}{}_{b}e^{b}\right)\,,
    \\
    & \nonumber \\
    \label{eq:deltaxiomegaab}
    \delta_{\xi}\omega^{ab}
    & =
    -\left(\imath_{\xi}R^{a}+\mathcal{D}  P_{\xi}{}^{ab}\right\}\,,
  \end{align}
\end{subequations}

\noindent
which are Lorentz-covariant and vanish unambiguously for $\xi=k$ on account of
the Killing vector equation and Eq.~(\ref{eq:Lorentzmomentummapeq}),
respectively.

Let us consider the KR field strength. Since it is a gauge-invariant 3-form
and, assuming $k$ is a Killing vector

\begin{equation}
  0
  =
  \delta_{k}H
  =
  -\mathcal{L}_{k}H
  =
  -(\imath_{k}d+d\imath_{k})H
  =
  -d\imath_{k}H\,,
\end{equation}

\noindent
on account of the Bianchi identity $dH=0$. This implies that 

\begin{equation}
  \imath_{k}H
  =
  -dP_{B\, k}^{(1)} +\mathfrak{h}^{(2)}\,,
\end{equation}

\noindent
where $\mathfrak{h}^{(2)}$ is a closed but not exact 2-form and
$P_{B\, k}^{(1)}$ is the KR momentum map 1-form, defined up to the derivative
of a $0$-form.

For the KR 2-form potential we have

\begin{equation}
  \begin{aligned}
    0
    & =
    \delta_{k}B
    =
    -\mathcal{L}_{k}B +\delta_{\Lambda^{(1)}_{B\, k}}B
    =
      -\imath_{k}H- d\left(\imath_{k}B -\Lambda^{(1)}_{B\, k}\right)
    \\
    & \\
    & =
    -\mathfrak{h}^{(2)}
    -d\left(\imath_{k}B-P_{B\, k}^{(1)} -\Lambda^{(1)}_{B\, k}\right)\,,
  \end{aligned}
\end{equation}

\noindent
which implies the independent vanishing of the exact and not exact
parts\footnote{There is another possibility: we could have assumed a weaker
  symmetry condition $\delta_{k}B=-\mathfrak{h}^{(2)}$, instead of
  $\delta_{k}B=0$. This weaker condition can be understood as the statement
  that $B$ is invariant under the GCT generated by $k$ up to a global
  higher-form transformation \cite{Barbagallo:2025qdy}. We will not consider
  this possibility here, although it is certainly a consistent one because the
  action would be invariant under those global 2-form symmetries.} and the
identification of the parameter of the KR compensating gauge
transformation\footnote{This is true up to the addition of an exact 1-form
  that can be absorbed in the ambiguity in the definition of
  $P_{B\, k}^{(1)}$.}

\begin{equation}
\label{eq:Lambda01Bk}
  \Lambda^{(1)}_{B\, k}
  =
  \imath_{k}B-P_{B\, k}^{(1)}\,.
\end{equation}

Then, using this parameter for a general $\xi$

\begin{equation}
\label{eq:deltaxiB}
  \delta_{\xi}B
  =
  -\mathcal{L}_{\xi}B +\delta_{\Lambda^{(1)}_{B\, \xi}}B
  =
  -\imath_{\xi}H -d\imath_{\xi}B +d(\imath_{\xi}B-P_{B\, \xi}^{(1)})
  =
  -\left(\imath_{\xi}H+dP_{B\, \xi}^{(1)} \right)\,,
\end{equation}

\noindent
which vanishes identically and unambiguously for $\xi=k$ by virtue of the
\textit{KR momentum-map equation} \cite{Elgood:2020mdx}

\begin{equation}
  \label{eq:KRmomentummapeq}
  \imath_{k}H+dP_{B\, k}^{(1)}=0\,.
\end{equation}

Let us now consider the RR field strengths. $G^{(2)}$ is an ordinary Maxwell
field strength with no Chern--Simons term and, following the same arguments
as with the KR field strength, we arrive at the following form of the parameter
of the compensating gauge transformation and of the transformation:\footnote{A
  closed but not exact 1-form could be added, related to a global 1-form
  symmetry, but we will not study that possibility here.}

\begin{subequations}
  \begin{align}
    \label{eq:Lambda0k}
    \Lambda_{\xi}^{(0)}
    & =
      \imath_{\xi}C^{(1)}-P_{\xi}^{(0)}\,,
    \\
    & \nonumber \\
    \delta_{\xi}C^{(1)}
    & =
      -\left(\imath_{\xi}G^{(2)}+dP_{\xi}^{(0)}\right)\,,
  \end{align}
\end{subequations}

\noindent
where, when $\xi=k$, the RR momentum map $0$-form $P_{k}^{(0)}$ is defined, up
to an additive constant, by the momentum map equation

\begin{equation}
  \label{eq:RR0momentummapeq}
\imath_{k}G^{(2)}+dP_{k}^{(0)}=0\,.
\end{equation}

Next, using the previous results

\begin{equation}
  \begin{aligned}
    0
    & =
    \delta_{k}G^{(4)}
    =
    -\mathcal{L}_{k}G^{(4)}
    =
    -(\imath_{k}d+d\imath_{k})G^{(4)}
    =
      -\imath_{k}(H\wedge G^{(2)}) -d\imath_{k}G^{(4)}
    \\
    & \\
    & =
    dP_{B\, k}^{(1)}\wedge G^{(2)}
    -H\wedge dP_{k}^{(1)}
      -d\imath_{k}G^{(4)}
      =
    d\left(P_{B\, k}^{(1)}\wedge G^{(2)}+HP_{k}^{(0)}-\imath_{k}G^{(4)}\right)\,,
  \end{aligned}
\end{equation}

\noindent
which implies the existence of a RR momentum map 3-form $P_{k}^{(3)}$
satisfying the momentum map equation\footnote{Here, as we have done before,
  and for similar reasons, we are ignoring the possibility of adding to this
  equation a closed but not exact 4-form.}

\begin{equation}
  \label{eq:RR2momentummapeq}
\imath_{k}G^{(4)}- P_{B\, k}^{(1)}\wedge G^{(2)} +dP_{k}^{(2)} -HP_{k}^{(0)} =0\,.
\end{equation}

The transformation of the 3-form under GCTs is, now

\begin{equation}
  \begin{aligned}
    0
    & =
    \delta_{k}C^{(3)}
    =
    -\mathcal{L}_{k}C^{(3)} +\delta_{\Lambda_{k}}C^{(3)}
    \\
    & \\
    & =
    -\left[\imath_{k}G^{(4)}
    -P_{B\, k}^{(1)}\wedge G^{(2)}
      -d\left( \Lambda^{(2)}_{k}-\imath_{k}C^{(3)}+P_{B\, k}^{(1)}\wedge
      C^{(1)}\right)
      -HP^{(0)}_{k} \right]\,.
  \end{aligned}
\end{equation}

\noindent
Using the momentum map equation (\ref{eq:RR2momentummapeq}), we can identify
the compensating gauge transformation parameter 

\begin{equation}
\label{eq:Lambda2k}
  \Lambda^{(2)}_{k}
  =
  \imath_{k}C^{(3)}-P_{k}^{(2)} -P_{B\, k}^{(1)}\wedge C^{(1)}\,,
\end{equation}

\noindent
and, now,

\begin{equation}
\begin{aligned}
    \delta_{\xi}C^{(3)}
    & =
      -\mathcal{L}_{\xi}C^{(3)} +\delta_{\Lambda_{\xi}}C^{(3)}
  \\
  & \\
    & =
      -\left(\imath_{\xi}G^{(4)}
      -P_{B\, \xi}^{(1)}\wedge G^{(2)}
      +dP_{\xi}^{(2)} 
      -HP_{\xi}^{(0)}\right)
            +\delta_{\xi}B\wedge C^{(1)}\,,
\end{aligned}
\end{equation}

\noindent
which vanishes identically for $\xi=k$ by virtue of the momentum map equation
(\ref{eq:RR2momentummapeq}) and because, by assumption and construction,
$\delta_{k}B=0$.

We can repeat the procedure for $G^{(6)}$ and $G^{(7)}$ but, since all these
field strengths have the same structure, using the RR momentum map polyform
$P_{\xi}=\sum_{n=0,2,4,6}P_{\xi}^{(n)}$ and the polyforms previously defined,
it is possible to show that the RR momentum maps satisfy the general equations

\begin{equation}
  \label{eq:RRgeneralmomentummapeq}
\imath_{k}G -P_{B\, k}^{(1)}\wedge G +d_{H^{-}}P_{k}=0\,,  
\end{equation}

\noindent
that the parameters of the RR compensating gauge transformations are

\begin{equation}
  \label{eq:Lambdaxi}
  \Lambda_{\xi}
  =
  \imath_{\xi}C-P_{\xi} -P_{B\, \xi}^{(1)}\wedge C\,,
\end{equation}

\noindent
and the transformation of the RR fields under GCTs are given by

\begin{equation}
  \label{eq:deltaxiC}
  \delta_{\xi}C
  =
  -\left(\imath_{\xi}G -P_{B\, \xi}^{(1)}\wedge G +d_{H^{-}}P_{\xi} \right)
  +\delta_{\xi}B\wedge C\,.
\end{equation}

Having defined the action of $\delta_{\xi}$ on all the fields, we can derive
the associated Noether charge with the usual algorithm: since $\delta_{\xi}$
is a combination of the standard Lie derivative and gauge transformations that
leave the action invariant up to a total derivative, it will also leave it
invariant up to a total derivative, which is nothing but

\begin{equation}
  \delta_{\xi}S
  =
  \int_{\mathcal{M}_{10}}\delta_{\xi}\mathbf{L}
  =
  \int_{\mathcal{M}_{10}}\left(-\mathcal{L}_{\xi}+\delta_{\Lambda_{\xi}}\right)\mathbf{L} 
  =
  -\int_{\mathcal{M}_{10}}d \left(\imath_{\xi}\mathbf{L}
    +d\mathbf{Y}^{(7)}\wedge \Lambda_{B\, \xi}^{(1)}\right)\,.
\end{equation}

On the other hand, from Eq.~(\ref{eq:generalvariation2a}) we get a different
expression for this variation. Substituting the results we have obtained above,
we get, after integration by parts and use of the Noether
identities,\footnote{We have checked they are identically satisfied, as
  expected, as a test of the consistency of this calculation but we will not
  give them here explicitly to simplify the exposition.} 

\begin{equation}
    \delta_{\xi} S
     =
      \int_{\mathcal{M}_{10}} d \left\{ \mathbf{\Theta}(\varphi,\delta_{\xi} \varphi)
      +\xi^{a}\mathbf{E}_{a}-P_{B\, \xi}^{(1)}\wedge
      (\mathbf{E}_{B}-C^{(1)}\wedge \mathbf{E}_{(3)})
      +P_{\xi}^{(0)}\mathbf{E}_{(1)}
      +P_{\xi}^{(2)}\wedge \mathbf{E}_{(3)}\right\}\,,
\end{equation}

\noindent
and, equating the two expressions we have obtained for $\delta_{\xi} S$ we
arrive at the off-shell conserved Noether current

\begin{equation}
  \label{eq:Jxi2A}
  \begin{aligned}
    \mathbf{J}[\xi]
    & =
    \mathbf{\Theta}(\varphi,\delta_{\xi} \varphi)
    +\xi^{a}\mathbf{E}_{a}-P_{B\, \xi}^{(1)}\wedge
    (\mathbf{E}_{B}-C^{(1)}\wedge \mathbf{E}_{(3)})
    +P_{\xi}^{(0)}\mathbf{E}_{(1)}
      +P_{\xi}^{(2)}\wedge \mathbf{E}_{(3)}
    \\
    & \\
    & \hspace{.5cm}
    +\imath_{\xi}\mathbf{L}
    +d\mathbf{Y}^{(7)}\wedge \Lambda_{B\, \xi}^{(1)}\,.
  \end{aligned}
\end{equation}

A long but straightforward calculation shows that this Noether current is
exact

\begin{equation}
  \mathbf{J}[\xi] =d\mathbf{Q}_{NW}[\xi]\,,
\end{equation}

\noindent
and that the Noether--Wald 8-form is given by

\begin{equation}
  \label{eq:QNW2a}
  \begin{aligned}
    \mathbf{Q}_{NW}[\xi]
    & =
    e^{-2\phi} \star (e^{a}\wedge e^{b})P_{\xi\, ab}
      +2\imath_{\xi}\star de^{-2\phi}
    \\
    & \\
    & \hspace{.5cm}
      -P_{B\, \xi}^{(1)}\wedge (e^{-2\phi}\star H)
    -P_{\xi}^{(0)}\mathbf{X}^{(8)}
    -P_{\xi}^{(2)}\mathbf{X}^{(6)}\,.
  \end{aligned}
\end{equation}

In general, the Noether--Wald charge is not conserved: its exterior derivative
is $\mathbf{J}[\xi]$ in Eq.~(\ref{eq:Jxi2A}), some of whose terms vanish
on-shell because they are proportional to the equations of motion and some of
whose terms (the presymplectic potential) vanish for $\xi=k$ because they are
linear in $\delta_{k}\varphi$. Under those conditions the surviving terms are just

\begin{equation}
  d\mathbf{Q}_{NW}[k]
  \doteq
  \imath_{k}\mathbf{L}
  +d\mathbf{Y}^{(7)}\wedge \Lambda_{B\, k}^{(1)}\,.
\end{equation}

The right-hand side of this equation must be a total derivative because the
left-hand side is and because, by assumption, $\delta_{k}$ annihilates all the
field of the theory and, therefore, $\delta_{k}\mathbf{L}=0$. 

In order to compute $\imath_{k}\mathbf{L}$ on-shell, we first compute
$\mathbf{L}$ on-shell. There is no unique way of doing this. The simplest
consists in using the dilaton equation

\begin{equation}
  \mathbf{L}
  =
  \mathbf{L}_{NSNS}+\mathbf{L}_{RR}
  =
  -2d\star d e^{-2\phi} -\tfrac{1}{2}\mathbf{E}_{\phi}+\mathbf{L}_{RR}
  \doteq
  -2d\star d e^{-2\phi}+\mathbf{L}_{RR}\,.
\end{equation}

Then,

\begin{equation}
  \label{eq:ikLonshell2A}
  \begin{aligned}
    \imath_{k} \mathbf{L}_{\rm on-shell}
    & =
      -2\imath_{k}d\star d e^{-2\phi}+\imath_{k}\mathbf{L}_{RR}
    \\
    & \\
    & =
      d\left( 2\imath_{k}\star d e^{-2\phi}\right)
      +\imath_{k}G^{(2)}\wedge \left(\tfrac{1}{2}\star G^{(2)}
      +\tfrac{1}{2}G^{(4)}\wedge B^{2}
     -\tfrac{1}{3}G^{(2)} \wedge B^{3}\right)
    \\
    & \\
    & \hspace{.5cm}
      +\imath_{k}G^{(4)}\wedge \left(\tfrac{1}{2}\star G^{(4)}
      -G^{(4)}\wedge B
      +\tfrac{1}{2}G^{(2)} \wedge B^{2}\right)
    \\
    & \\
    & \hspace{.5cm}
      +\tfrac{1}{2} G^{(2)} \wedge \imath_{k}\star G^{(2)}
      +\tfrac{1}{2} G^{(4)} \wedge \imath_{k}\star G^{(4)}
      -d\mathbf{Y}^{(7)}\wedge \imath_{k}B\,.
  \end{aligned}
\end{equation}

We are now going to use the RR momentum map equations
(\ref{eq:RR0momentummapeq}) and (\ref{eq:RR2momentummapeq}) to eliminate
$\imath_{k}G^{(2)}$ and $\imath_{k}G^{(4)}$ from the above expression. We
could also use the general momentum map equation
(\ref{eq:RRgeneralmomentummapeq}) to eliminate $\imath_{k}\star G^{(2)}$ and
$\imath_{k}\star G^{(4)}$ using the fact that they are equal to
$-\imath_{k}G^{(8)}$ and $\imath_{k}G^{(6)}$ respectively but, for later use,
it is convenient to derive a specific equation for the magnetic momentum maps
associated to the Hodge duals of the field strengths. Following the same steps
as before but using the equations of motion

\begin{equation}
  \label{eq:genericRReoms}
d_{H^{+}}\star G  \equiv (d+H\wedge ) \star G=0\,,
\end{equation}

\noindent
instead of the Bianchi identities

\begin{equation}
d_{H^{-}} G  \equiv (d-H\wedge ) G=0\,,
\end{equation}

\noindent
we find 

\begin{equation}
  0
  =
  \delta_{k}\star G
  =
  -d_{H^{+}}\left(\imath_{k}\star G+P_{B\, k}^{(1)}\wedge \star G\right)
\end{equation}

\noindent
which leads to the magnetic momentum map equation

\begin{equation}
  \label{eq:RRmagneticmomentummapequations}
\imath_{k}\star G+P_{B\, k}^{(1)}\wedge \star G +d_{H^{+}}\tilde{P}_{k} = 0\,.
\end{equation}

Of course, the magnetic momentum maps $\tilde{P}_{k}^{(n)}$ are directly
related to the electric ones $P_{k}^{(n)}$ because of the relations
Eq.~(\ref{eq:dualityconstraints}):

\begin{equation}
  \label{eq:dualityconstraintsRRmomentummaps}
  \tilde{P}_{k}^{(n)} =  (-1)^{[(8-n)/2]}  P_{k}^{(n)}\,.
\end{equation}

\noindent
However, in the democratic formulation that we will use in
Section~\ref{sec-democratic}, it is convenient to treat them as independent
objects until the very end of the calculation.

Going back to our calculation Eq.~(\ref{eq:ikLonshell2A}), using the duality
relations, the momentum map equations and the equations of motion and Bianchi
identities, we arrive to 

\begin{equation}
    \imath_{k} \mathbf{L}_{\rm on-shell}
     =
      d\mathbf{w}_{k} -d\mathbf{Y}^{(7)}\wedge \Lambda_{B\, k}^{(1)}\,, 
\end{equation}

\noindent
where

\begin{equation}
  \begin{aligned}
    \mathbf{w}_{k}
    & =
    2\imath_{k}\star d e^{-2\phi}
    -P_{k}^{(0)}\left(\mathbf{X}^{(8)}-\tfrac{1}{2}\star G^{(2)}\right)
      -\tfrac{1}{2}P_{k}^{(4)}\wedge G^{(4)}
    \\
    & \\
    & \hspace{.5cm}
    -P_{k}^{(2)}\wedge \left(\mathbf{X}^{(6)}-\tfrac{1}{2}\star G^{(4)}\right)
    -\tfrac{1}{2}P_{k}^{(6)}\wedge G^{(2)}\,.
  \end{aligned}
\end{equation}

The on-shell conserved generalized Komar charge of the $\mathcal{N}=2A,d=10$
supergravity theory is, then,

\begin{equation}
  \label{eq:Komarcharge2A}
  \mathbf{K}[k]
  \equiv
  -\mathbf{Q}_{NW}[k]+\mathbf{w}_{k}
  =
    -e^{-2\phi} \star (e^{a}\wedge e^{b})P_{\xi\, ab}
    +P_{B\, \xi}^{(1)}\wedge (e^{-2\phi}\star H)
    +\tfrac{1}{2}P_{k}\wedge \star G\,.
\end{equation}

This charge is manifestly gauge-invariant because each of its terms
is.\footnote{Since the momentum maps are defined up to the addition of total
  derivatives, the generalized Komar charge has ambiguities that have to be
  studied separately. As shown in
  Refs.~\cite{Barbagallo:2025qdy,Ortin:2026fzm}, these ambiguities can have
  important effects like the modification of the magnetic momentum map
  equations and the loss of the conservation property of the generalized Komar
  charge, which may get a source term. This is what happens in
  $\mathcal{N}=1,d=5$ supergravity coupled to vector multiplets and it
  explains the existence of massive, globally regular, charged,
  asymptotically-flat black-hole microstate geometries (\textit{fuzzballs})
  \cite{Gibbons:2013tqa}.  } However, those terms do not have the structure
one needs to derive Smarr formulas or, simply, to be easily interpreted in
terms of physically meaningful objects like potentials and conserved
charges. In all the cases examined so far, the terms involving gauge fields
can be rewritten in the form

\begin{equation}
  \Phi_{k}^{(n)}\wedge \mathbf{F}^{(8-n)}\,,
\end{equation}

\noindent
where $\mathbf{F}^{(8-n)}$ is an on-shell closed combination of field
strengths and potentials and $\Phi_{k}^{(n)}$ is an $n$-form potential built
out of momentum maps and fields which is closed on-shell when $k=0$. This last
property allows us to prove generalized, restricted, 0\textsuperscript{th}
laws \cite{Elgood:2020svt}.

In Section~\ref{sec-n2ad10gauge} we have constructed charges which are
conserved on-shell for the KR and RR fields (Eqs.~(\ref{eq:KRcharge2A}) and
(\ref{eq:RRcharges2A}), respectively) which have the above structure.  In
those definitions, the role of $\mathbf{F}^{(8-n)}$ is played by the so-called
Page charges. In order to rewrite Eq.~(\ref{eq:Komarcharge2A}) in terms of
them we first need to construct Page charges using the Hodge duals of the
field strengths.

Since

\begin{equation}
  d\left(e^{B}\wedge \star G\right)
  =
  e^{B}d_{H^{+}}\star G
  =
  0\,,
\end{equation}

\noindent
we conclude that

\begin{equation}
  \label{eq:tildeSdefinition}
  \widetilde{\star G}
  \equiv
  e^{B}\wedge \star G
  \equiv
  d\tilde{S}\,,
\end{equation}

\noindent
for some potentials $\tilde{S}$ and can be used as Page
charges.\footnote{\label{foot:dualthings}Using the duality relations
  Eq.~(\ref{eq:dualityconstraints}) one can show that they also hold between
  the Page charges:
  \begin{equation}
\label{eq:dualityconstraintsPagecharges}
\tilde{G}^{(10-n)} = (-1)^{[n/2]} \widetilde{\star G^{(n)}}\,,
\end{equation}
and lead to the following relations between the \textit{Page potentials}
\begin{equation}
\label{eq:dualityconstraintsPagepotentials}
S^{(n)} = (-1)^{[(9-n)/2]} \tilde{S}^{(n)}\,.
\end{equation}
} Using them and the KR Page charge $\tilde{H}$ defined in
Eq.~(\ref{eq:tildeHdef}) we have

\begin{equation}
  \begin{aligned}
    \mathbf{K}[k]
    & =
      -e^{-2\phi} \star (e^{a}\wedge e^{b})P_{k\, ab}
      +P_{B\, k}^{(1)}\wedge \left[ \tilde{H} +\tfrac{1}{2} Ce^{-B}\wedge e^{B}\star G \right]
      +\tfrac{1}{2}P_{k}e^{-B}\wedge e^{B}\star G
    \\
    & \\
    & =
      -e^{-2\phi} \star (e^{a}\wedge e^{b})P_{k\, ab}
      +P_{B\, k}^{(1)}\wedge \tilde{H}
      +\tfrac{1}{2}\left(P_{k}+P_{B\, k}^{(1)}\wedge C\right)e^{-B}\wedge  \widetilde{\star G}\,.
  \end{aligned}
\end{equation}

Because of the KR momentum map equation (\ref{eq:KRmomentummapeq})

\begin{equation}
dP_{B\, k}^{(1)} \stackrel{k=0}{=} 0\,. 
\end{equation}

On the other hand,

\begin{equation}
  d\left\{\left(P_{k}+P_{B\, k}^{(1)}\wedge C\right)e^{-B}\right\}
  =
  e^{-B}d_{H^{-}}\left(P_{k}+P_{B\, k}^{(1)}\wedge C\right)
  =
  e^{-B} \left( -\imath_{k}G -\imath_{k}H\wedge C \right)
  \stackrel{k=0}{=} 0\,.
\end{equation}

Thus, we can define the potentials

\begin{equation}
  \label{eq:potentialsdef}
  \Phi_{B\, k}^{(1)} \equiv  P_{B\, k}^{(1)}\,,
  \hspace{1cm}
    \Phi_{k} \equiv  \left(P_{k}+P_{B\, k}^{(1)}\wedge C\right)e^{-B}\,,
\end{equation}

\noindent
and rewrite the generalized Komar charge in the equivalent but physically
meaningful form

\begin{equation}
  \label{eq:Komarcharge2APage}
    \mathbf{K}[k]
    =
      -e^{-2\phi} \star (e^{a}\wedge e^{b})P_{k\, ab}
      +\Phi_{B\, k}^{(1)}\wedge \tilde{H}
      +\tfrac{1}{2}\Phi_{k}\wedge  \widetilde{\star G}\,.
\end{equation}

Now, none of the terms of this charge, except for the first one, which is
the standard Komar charge of pure GR

\begin{equation}
  e^{-2\phi} \star (e^{a}\wedge e^{b})P_{k\, ab}
  =
  e^{-2\phi} \star d\hat{k}\,,
  \hspace{1cm}
  \hat{k}
  =
  k_{\mu}dx^{\mu}\,,
\end{equation}

\noindent
is gauge invariant, although we have shown that the whole Komar charge is.

\section{The massive $\mathcal{N}=2A,d=10$ supergravity}
\label{sec-n2ad10msugra}

Romans' massive $\mathcal{N}=2A,d=10$ supergravity theory \cite{Romans:1985tz}
has the same field content as the massless one, but new mass- ($m$-) dependent
interactions which occur 

\begin{enumerate}
\item Implicitly through modified RR field strengths

\begin{equation}
  G^{(2)} \equiv dC^{(1)}+mB\,,
  \hspace{1cm}
  G^{(4)} \equiv dC^{(3)}-H\wedge C^{(1)}+\tfrac{1}{2}mB^{2}\,,
\end{equation}

\noindent
invariant under the gauge transformations

\begin{equation}
  \label{eq:gaugetransformations2Am}
  \delta_{\Lambda} C^{(1)} = d\Lambda^{(0)}-m\Lambda_{B}^{(1)}\,,   
  \hspace{1cm}
  \delta_{\Lambda} C^{(3)} = d\Lambda^{(2)}-H\Lambda^{(0)}-mB\wedge \Lambda_{B}^{(1)} \,.
\end{equation}

\item Explicitly through  $m$-dependent terms in the RR part of the action
Eq.~(\ref{eq:n2ad10action})

\begin{equation}
  \label{eq:LRR2Am}
  \begin{aligned}
    \mathbf{L}_{RR}
    & =
      \tfrac{1}{2}\star m^{2}
      +\tfrac{1}{2}G^{(2)}\wedge \star G^{(2)}
      +\tfrac{1}{2}G^{(4)}\wedge \star G^{(4)}
    \\
    &  \\
    & \hspace{.5cm}
      -\tfrac{1}{2}G^{(4)}\wedge G^{(4)}\wedge B
      +\tfrac{1}{2}G^{(4)}\wedge G^{(2)} \wedge B^{2}
      -\tfrac{1}{3!}G^{(2)}\wedge G^{(2)} \wedge B^{3}
    \\
    &  \\
    & \hspace{.5cm}
      -\tfrac{1}{3!}mG^{(4)}\wedge B^{3}
      +\tfrac{1}{8}mG^{(2)}\wedge B^{4}
      -\tfrac{3}{5!}m^{5}B^{5}\,.
  \end{aligned}
\end{equation}
\end{enumerate}

Again, written in this form in which the RR potentials only occur through the
RR field strengths, the action only changes by a total derivative under the
gauge transformations of $B$:
  
\begin{equation}
  \label{eq:deltaLambdaS2Am}
  \begin{aligned}
    \delta_{\Lambda}S
    & =
      -\int_{\mathcal{M}_{10}} d\mathbf{Y}^{(7)}\wedge \delta_{\Lambda}B\,,
    \\
    & \\
    \mathbf{Y}^{(7)}
    & \equiv
      \tfrac{1}{2}\left[\left(G^{(4)}-\tfrac{1}{2}mB^{2}\right)
      -B\wedge \left(G^{(2)}-mB\right) \right] \wedge (C^{(3)}-B\wedge C^{(1)})\,.
    \\
    & \\
    d\mathbf{Y}^{(7)}
    & \equiv
      \tfrac{1}{2}\left[\left(G^{(4)}-\tfrac{1}{2}mB^{2}\right)
      -B\wedge \left(G^{(2)}-mB\right) \right]
      \wedge \left[\left(G^{(4)}-\tfrac{1}{2}mB^{2}\right)
      -B\wedge \left(G^{(2)}-mB\right) \right]\,.
  \end{aligned}
\end{equation}

The only two equations of motion in Eq.~(\ref{eq:eom2A}) which have explicit
massive modifications (beyond the implicit dependence of the RR field strengths
and $\mathbf{L}_{RR}$ on $m$) are

\begin{subequations}
  \begin{align}
    \mathbf{E}_{a}
    & =
       \mathbf{E}_{a}^{m=0}+\tfrac{1}{2}\imath_{a}\star m^{2}\,,
      \\
    & \nonumber \\
    \mathbf{E}_{B}
    & =
    \mathbf{E}_{B}^{m=0} +m\star G^{(2)}\,,
  \end{align}
\end{subequations}

\noindent
and the presymplectic potential is still given by Eq.~(\ref{eq:Theta2A}), but
now  $\mathbf{X}^{(8)}$ and $\mathbf{X}^{(6)}$ get explicit massive corrections

\begin{subequations}
  \begin{align}
    \mathbf{X}^{(8)}
    &  =
      \mathbf{X}^{(8)\, m=0}
      +\tfrac{1}{8}mB^{4}\,,
    \\
    & \nonumber \\
    \mathbf{X}^{(6)}
    & =
      \mathbf{X}^{(6)\, m=0}
      -\tfrac{1}{3!}mB^{3}\,,
  \end{align}
\end{subequations}

\noindent
but Eqs.~(\ref{eq:X6relation}) (\ref{eq:X8relation}) are still satisfied,
while Eq.~(\ref{eq:ZYrelation}) gets a correction

\begin{equation}
    \label{eq:ZYrelationmassive}
    d\left(\mathbf{Z}^{(7)} +\mathbf{Y}^{(7)}\right)
     =
      -\mathbf{E}_{B}+ m\left(\star G^{(2)} +\star G^{(4)}\wedge B
        -\tfrac{1}{2}G^{(4)}\wedge B^{2}+\tfrac{1}{3!}G^{(2)}\wedge B^{3}
      -\tfrac{1}{4!}mB^{4}\right)\,.  
\end{equation}

The Bianchi identity of $G^{(4)}$ has the same form as in the massless case
but that of $G^{(2)}$ does not:

\begin{equation}
dG^{(2)}-mH=0\,.
\end{equation}

This suggests to introduce formally a RR $0$-form field strength which should be
constant and equal to $m$

\begin{equation}
G^{(0)} =m\,,  
\end{equation}

\noindent
so that the form of the Bianchi identity is similar to that of the rest of RR
field strengths: that of $G^{(4)}$ can be calculated using the
definition. Those of $G^{(6)}$ and $G^{8}$ are the equations of motion of
$C^{(1)}$ and $C^{(3)}$ and solving them we find the general form of the RR
field strengths in the massive $\mathcal{N}=2A,d=10$ supergravity theory

\begin{equation}
  \label{eq:generalRRfieldstrengths}
G = d_{H^{-}}C +me^{B}\,,  
\end{equation}

\noindent
and their Bianchi identities

\begin{equation}
  \label{eq:generalRRBianchiidentities}
d_{H^{-}}G = 0\,,  
\end{equation}

\noindent
and equations of motion

\begin{equation}
    \label{eq:generalRReoms}
d_{H^{+}}\star G = 0\,,  
\end{equation}

\noindent
The general form of the gauge transformations of the RR fields is, now,

\begin{equation}
  \label{eq:generalRRgaugetransformations}
  \delta_{\Lambda}C
  =
  d_{H^{-}}\Lambda -m e^{B}\wedge \Lambda_{B}^{(1)}\,.
\end{equation}

The set of RR fields can be further extended to include a RR 9-form potential,
$C^{(9)}$, whose 10-form field strength, $G^{(10)}$, given by the above
general expression, is dual to $G^{(0)}$ according to the general formula
Eq.~(\ref{eq:dualityconstraints}), that is: $G^{(0)}=-\star G^{(10)}$. Thus,
in this section, from now on,

\begin{subequations}
  \begin{align}
    C & \equiv \sum_{n=1,3,5,7,9}C^{(n)}\,,
    \\
    & \nonumber \\
    G & \equiv \sum_{n=0,2,4,6,8,10}G^{(n)}= d_{H^{-}}C+me^{B}\,.
  \end{align}
\end{subequations}

With these definitions, the RR Page charges and their potentials are now

\begin{subequations}
  \begin{align}
    \label{eq:Sdefinitionmassive}
    \tilde{G}
    & \equiv
    e^{-B}\wedge G
    =
    d(e^{-B}\wedge C)+m
    \equiv
    dS +m\,,
    \\
    & \nonumber \\
    \label{eq:tildeSdefinitionmassive}
    \widetilde{\star G}
    & \equiv
    e^{B}\wedge \star G
    \equiv
    d\tilde{S}-m\,.
  \end{align}
\end{subequations}

The duality relations between Page charges and potentials
Eqs.~(\ref{eq:dualityconstraintsPagecharges}) and
(\ref{eq:dualityconstraintsPagepotentials}) are satisfied now for all the
relevant values of $n$.

Finally, under gauge  transformations

\begin{subequations}
  \label{eq:Pagechargesgaugetransformationsmassive}
  \begin{align}
    \delta_{\Lambda} S
    & =
      d\left[ (\Lambda -\Lambda_{B}^{(1)}\wedge C)\wedge e^{-B}\right]
      -\Lambda_{B}^{(1)}\wedge \tilde{G}\,,
    \\
    & \nonumber \\
    \delta_{\Lambda} \tilde{G}
    & =
      -d\left(\Lambda_{B}^{(1)}\wedge \tilde{G}\right)\,.
  \end{align}
\end{subequations}

\subsection{Symmetries and conserved charges}
\label{sec-n2amd10sugrasymmetriesandcharges}

\subsubsection{Local Lorentz transformations}
\label{sec-n2amd10LL}

The situation of the massive theory with respect to local Lorentz
transformations is exactly the same as in the massless case studied in
Section~\ref{sec-n2ad10LL}, to which we refer the reader.

\subsubsection{NSNS and RR gauge transformations}
\label{sec-n2amd10gauge}

The KR gauge transformations of this theory have a completely different
character: they act as shifts on $C^{(1)}$ 

\begin{equation}
  \delta_{\Lambda_{B}^{(1)}}C^{(1)}
  =
  -m\Lambda_{B}^{(1)}\,,
\end{equation}

\noindent
and, consequently, they could be used to eliminate it. If we do that,

\begin{equation}
G^{(2)} = mB\,,   
\end{equation}

\noindent
and its kinetic term becomes a mass term for the KR 2-form whose gauge
symmetry is broken. In other words: $C^{(1)}$ is just a St\"uckelberg field
whose only role is to formally preserve the KR gauge symmetry. In principle,
we do not expect a conserved charge associated to either $B$ or $C^{(1)}$, but
it is worth studying the situation in detail, applying first the same
algorithm we used in the massless case, Section~\ref{sec-n2ad10gauge}.

Under a combined gauge transformation, the action transforms as

\begin{equation}
  \begin{aligned}
    \delta_{\Lambda} S
    & =
    \int_{\mathcal{M}_{10}} \left\{
    -\left[d\mathbf{E}_{B}+m\left(\mathbf{E}_{(1)}+\mathbf{E}_{(3)}\wedge
      B\right)\right]
      \wedge \Lambda^{(1)}_{B}
            \right.
    \\
    & \\
    & \hspace{.5cm}
      \left.
    +\left(d\mathbf{E}_{(1)}-\mathbf{E}_{(3)}\wedge H\right)\Lambda^{(0)}
      +d\mathbf{E}_{(3)}\wedge \Lambda^{(2)}
      \right.
    \\
    & \\
    & \hspace{.5cm}
      \left.
      +d\left[\mathbf{\Theta}(\varphi,\delta_{\Lambda} \varphi)
      +\mathbf{E}_{B}\wedge \Lambda^{(1)}_{B}
      -\mathbf{E}_{(1)}\Lambda^{(0)}
      -\mathbf{E}_{(3)}\wedge \Lambda^{(2)}\right]
      \right\}\,.
  \end{aligned}
\end{equation}

Each of the bulk terms proportional to a gauge parameter vanishes identically
(Noether identities) and $\delta_{\Lambda}S$ is a total derivative

\begin{equation}
    \delta_{\Lambda} S
     =
    \int_{\mathcal{M}_{10}} d\left[\mathbf{\Theta}(\varphi,\delta_{\Lambda} \varphi)
      +\mathbf{E}_{B}\wedge \Lambda^{(1)}_{B}
      -\mathbf{E}_{(1)}\Lambda^{(0)}
      -\mathbf{E}_{(3)}\wedge \Lambda^{(2)}\right]\,,
\end{equation}

\noindent
which should be identical to the one computed in
Eq.~(\ref{eq:deltaLambdaS2Am}). It follows that the current

\begin{equation}
  \mathbf{J}[\Lambda]
  =
  \mathbf{\Theta}(\varphi,\delta_{\Lambda} \varphi)
      +\mathbf{E}_{B}\wedge \Lambda^{(1)}_{B}
      -\mathbf{E}_{(1)}\Lambda^{(0)}
      -\mathbf{E}_{(3)}\wedge \Lambda^{(2)}
      +\mathbf{Y}^{(7)}\wedge d\Lambda^{(1)}_{B}\,,
\end{equation}

\noindent
is identically conserved and should be exact and, following the same steps as
in the massless, we arrive to exactly the same result Eq.~(\ref{eq:QLambda2a})
although, now, the forms involved in the expression of the charge have
additional, $m$-dependent terms.

The additional $m$-dependent terms in $\mathbf{X}^{(6)}$ and
$\mathbf{X}^{(8)}$ are such that we still can relate them to the Page charges
defined in Eq.~(\ref{eq:Sdefinitionmassive}).  We find the relations

\begin{subequations}
  \begin{align}
    \mathbf{X}^{(6)}
    & =
      G^{(6)} -G^{(4)}\wedge B +\tfrac{1}{2} G^{(2)}\wedge B^{2}
      -\tfrac{1}{3!}G^{(0)}B^{3}
      =
      \left.\tilde{G}\right|_{6}\,,
    \\
    & \nonumber \\
    \mathbf{X}^{(8)}+\mathbf{X}^{(6)}\wedge B
    & =
      -G^{(8)} +G^{(6)}\wedge B -\tfrac{1}{2}G^{(4)}\wedge B^{2}
      +\tfrac{1}{3!}G^{(2)}\wedge B^{3} 
      -\tfrac{1}{4!}G^{(0)}B^{4}
      =
      -\left.\tilde{G}\right|_{8}\,,
  \end{align}
\end{subequations}

\noindent
which confirm that, again, the RR charges will be constructed with the Page charges.

For the KR term, using the RR Page charges and potentials $\tilde{S}$ defined
in Eqs.~(\ref{eq:Sdefinitionmassive}) and (\ref{eq:tildeSdefinitionmassive}),
we find, on-shell and up to total derivatives 

\begin{equation}
    \mathbf{Z}^{(7)}+\mathbf{Y}^{(7)}
     =
      e^{-2\phi}\star H -\tfrac{1}{2}\star G\wedge C
      -\tfrac{1}{2}mS^{(7)}\,,
\end{equation}

\noindent
and we can define the KR Page charge

\begin{equation}
\label{eq:KRPagechargemassive}
    \tilde{H}
    \equiv
      e^{-2\phi}\star H -\tfrac{1}{2}\star G\wedge C
      -\tfrac{1}{2}m\tilde{S}^{(7)}\,,  
\end{equation}

\noindent
where now a term $\star G^{(10)} \wedge C^{(7)}$ is present in the product of
polyforms and and where $\tilde{S}^{(7)}$ is the rank-7 term in the polyform
$\tilde{S}$. This term indicates that, in this theory, $B^{(6)}$, whose field
strength will now be

\begin{equation}
  H^{(7)}
  =
  e^{-2\phi}\star H
  = dB^{(6)}+\tfrac{1}{2}\star G\wedge C+\tfrac{1}{2}m\tilde{S}^{(7)}\,,
\end{equation}

\noindent
is just a St\"uckelberg field that gives mass to $\tilde{S}^{(7)}$.

Thus, contrary to our initial expectations, we can construct RR and KR
conserved charges as in Eqs.~(\ref{eq:RRcharges2A}) and (\ref{eq:KRcharge2A})
because we have been able to construct Page charges for all of them. The same
is true in the Proca theory

\begin{equation}
  S[A]
  =
  \int_{\mathcal{M}_{4}}
  \left\{\tfrac{1}{2}F\wedge \star F +\tfrac{1}{2}m^{2}A\wedge \star A \right\}\,. 
\end{equation}

\noindent
If we introduce a St\"uckelberg field $\phi$

\begin{equation}
  S[A,\phi]
  =
  \int_{\mathcal{M}_{4}}
  \left\{\tfrac{1}{2}F\wedge \star F +\tfrac{1}{2}\left(d\phi+mA\right)\wedge \star
  \left(d\phi+mA\right) \right\}\,, 
\end{equation}

\noindent
that formally restores the gauge invariance

\begin{equation}
  \delta_{\Lambda}A = d\Lambda\,,
  \hspace{1cm}
  \delta_{\Lambda}\phi = -m\Lambda\,,  
\end{equation}

\noindent
and dualize it into a 2-form $B$\,,

\begin{equation}
d\phi+mA \equiv \star dB \equiv \star H\,,  
\end{equation}

\noindent
we obtain the action

\begin{equation}
  S[A,B]
  =
  \int_{\mathcal{M}_{4}}
  \left\{\tfrac{1}{2}F\wedge \star F -\tfrac{1}{2}H\wedge \star H +mB\wedge F\right\}\,, 
\end{equation}

\noindent
which is invariant under the gauge transformations

\begin{equation}
  \delta_{\Lambda}A = d\Lambda^{(0)}\,,
  \hspace{1cm}
  \delta_{\Lambda}B = d\Lambda^{(1)}\,,  
\end{equation}

\noindent
and for which we can formally construct the 2-form charges

\begin{subequations}
  \begin{align}
    \mathbf{Q}
    & =
      \star F +mB\,,
    \\
    & \nonumber \\
    \mathbf{Q}[\Lambda^{(1)}]
    & =
      \Lambda^{(1)}\wedge (\star H -mA)\,,
  \end{align}
\end{subequations}

\noindent
which are conserved on-shell when $d\Lambda^{(1)}=0$. We may also use $F$ as
magnetic 2-form charge but, since the equations of motion need $B$, in
principle we cannot use $H$ as a magnetic 3-form charge.

The Maxwell field in the last version of the theory can be dualized, leading
to

\begin{equation}
  S[\tilde{A},B]
  =
  \int_{\mathcal{M}_{4}}
  \left\{\tfrac{1}{2}\left(d\tilde{A}+mB\right)\wedge \star \left(d\tilde{A}+mB\right) -\tfrac{1}{2}H\wedge \star H\right\}\,, 
\end{equation}

\noindent
which shows that $B$ is a massive field and $\tilde{A}$ is its St\"uckleberg
field.  The role of the massive and St\"uckelberg fields is reversed by
dualization ($B$ and $C^{(1)}$ versus $S^{(7)}$ and $B^{(6)}$).

\subsubsection{General coordinate transformations}
\label{sec-n2amd10gct}

The transformations $\delta_{\xi}$ acting on the NSNS fields have exactly the
same form as in the massless case, and they are given in
Eqs.~(\ref{eq:deltaxiea}), (\ref{eq:deltaxiomegaab}) and (\ref{eq:deltaxiB}),
with the Lorentz and KR momentum maps determined by
Eqs.~(\ref{eq:Lorentzmomentummapeq}) and (\ref{eq:KRmomentummapeq}),
respectively, and with the parameters of the Lorentz and KR compensating
transformations given by Eqs.~(\ref{eq:sigmaabk}) and
(\ref{eq:Lambda01Bk}). The generic solution to the Lorentz momentum map
equation is the Killing bivector Eq.~(\ref{eq:Killingbivector}). On the other
hand, since the RR gauge transformations and field strengths are modified by
$m$-dependent terms, we have to study the derivation of the RR momentum map
equations again.

It turns out that there are no explicit $m$-dependent terms neither in the
transformations of the potentials nor in the parameters of the compensating RR
gauge transformations or in the momentum map equations which keep the general
form given in Eqs.~(\ref{eq:deltaxiC}), (\ref{eq:Lambdaxi}) and
(\ref{eq:RRgeneralmomentummapeq}), respectively. Observe, though, that since
the polyform $G$ now includes $G^{(0)}=m$, the rank-1 term in the momentum map
polyform equation that defines $P_{k}^{(0)}$ is, in fact, modified by an
$m$-dependent term

\begin{equation}
  \label{eq:RR1formmomentummapeq}
\imath_{k}G^{(2)} -mP_{B\, k}^{(1)} +d P_{k}^{(0)}=0\,.
\end{equation}

The algorithm takes us, then, to a Noether--Wald 8-form which has the same form
as in the massless case, Eq.~(\ref{eq:QNW2a}) but with implicit $m$-dependence
in $\mathbf{X}^{(6)}$ and $\mathbf{X}^{(8)}$.

In order to find the generalized Komar charge we have to compute
$\imath_{k}\mathbf{L}_{on-shell}$, as in the massless case. In the calculation
we have to use the electric RR momentum map equations as well as the magnetic
ones Eq.~(\ref{eq:RRmagneticmomentummapequations}). Furthermore Proceeding in
exactly in the same way as in the massless case, we find that there is a term

\begin{equation}
\tfrac{1}{2}\imath_{k}\star m^{2} = -\tfrac{1}{2}m^{2}\star \hat{k}\,,  
\end{equation}

\noindent
which, evidently, was not present in the massless case and which does not
immediately look like a total derivative. However, since $d\star\hat{k}$ is
the trace of the Killing vector equation, it must vanish and, locally,
following Refs.~\cite{Magnon:1985sc,Bazanski:1990qd,Kastor:2008xb} we can
write

\begin{equation}
  \label{eq:R8def}
 m\star\hat{k} = d R_{k}^{(8)}\,, 
\end{equation}

\noindent
for some 8-form $R_{k}^{(8)}$. This leads to

\begin{equation}
      \imath_{k}\mathbf{L}_{on-shell} 
 =
    d\left\{\cdots -\tfrac{1}{2} G^{(0)}R^{(8)} \right\}
    -\tfrac{1}{2}m \left(P_{k}^{(6)}\wedge H -P_{B\, k}^{(1)}\wedge \star G^{(2)}\right)\,.
  \end{equation}

The last, $m$-dependent, term does not have the form of a total derivative but
it is on-shell closed and we can express it locally as a total derivative. The
correct way of doing this is to realize that
$m\star\hat{k} = -\imath_{k}G^{(10)}$ and, using the general RR momentum-map
equation (\ref{eq:RRgeneralmomentummapeq}) we should rewrite this term in the form

\begin{equation}
  m\star\hat{k}
  =
  -\imath_{k}G^{(10)}
  =
  -P_{B\, k}^{(1)}\wedge G^{(8)} +dP_{k}^{(8)}- H\wedge P_{k}^{(6)}\,,
\end{equation}

\noindent
\textit{i.e.}~as the sum of an exact and an on-shell closed but not exact form
missing in Eq.~(\ref{eq:R8def}). Thus,\footnote{A more systematic way of doing
  this is to promote $G^{(0)}=m$ to a scalar field in the action, adding at
  the same time a Lagrange multiplier that enforces the constraint $dm=0$
  on-shell, as done in
  Refs.~\cite{Bergshoeff:1996ui,Bergshoeff:2001pv,Meessen:2022hcg}, following
  Refs.~\cite{Aurilia:1980xj,Duff:1980qv}.}

\begin{equation}
 H\wedge P_{k}^{(6)} -P_{B\, k}^{(1)}\wedge \star G^{(2)}
  =
  d\left(P_{k}^{(8)}-R_{k}^{(8)}\right)\,,
\end{equation}

\noindent
and our result simply has one more term with respect to the results of the
massless case:

\begin{equation}
      \imath_{k}\mathbf{L}_{on-shell} 
  =
  d\left\{\cdots -\tfrac{1}{2} P_{k}^{(8)} G^{(0)} \right\}\,.
\end{equation}

The generalized Komar charge of Roman's, massive, $\mathcal{N}=2A,d=10$
supergravity has exactly the same form as in the massless case,
Eq.~(\ref{eq:Komarcharge2A}) but, since the polyform $G$ contains on more term
($G^{(0)}=m$) it actually has one more term, apart from the $m$-dependence
implicit in the RR field strengths. It is closed on-shell by construction.

Again, we would like to rewrite it in terms of potentials and Page-type
charges. The potentials are defined in Eqs.~(\ref{eq:potentialsdef}) but, now,

\begin{equation}
  \label{eq:dphi0}
d\Phi_{k}^{(0)}= dP_{k}^{(0)} \stackrel{k=0}{=} mP_{B\, k}^{(1)}\,,
\end{equation}

\noindent
\textit{i.e.}~$\Phi_{k}^{(0)}$ does not satisfy the property that leads to the
generalized zeroth law in black-hole thermodynamics.

Proceeding as in the massless case, we get

\begin{equation}
   \label{eq:Komarcharge2AmassivePage}
  \mathbf{K}[k]
  =
  -e^{-2\phi} \star (e^{a}\wedge e^{b})P_{k\, ab}
  +P_{B\, k}^{(1)}\wedge \tilde{H}
  +\tfrac{1}{2}\Phi_{k} \wedge  \widetilde{\star G}
  +\tfrac{1}{2}mP_{B\, k}^{(1)}\wedge \tilde{S}^{(7)}\,,
\end{equation}

\noindent
where the Page charges $\widetilde{\star G}$ have been defined in
Eq.~(\ref{eq:tildeSdefinition}) and (\ref{eq:tildeSdefinitionmassive}).

The last term in Eq.~(\ref{eq:Komarcharge2AmassivePage}) and the term
involving $\Phi_{k}^{(0)}$ do not fit in the pattern we have observed in all
the cases studied until now. However, using
Eqs.~(\ref{eq:tildeSdefinitionmassive}) and
(\ref{eq:dualityconstraintsPagepotentials}) and Eq.~(\ref{eq:dphi0}), we can
rewrite these two terms as a total derivative

\begin{equation}
  \begin{aligned}
    \Phi_{k}^{(0)}\left.\widetilde{\star G} \right|_{8}
    -mP_{B\, k}^{(1)}\wedge S^{(7)}
    & =
    -\Phi_{k}^{(0)}dS^{(7)} -mP_{B\, k}^{(1)}\wedge S^{(7)}
    \\
    & \\
    & = d\left(-\Phi_{k}^{(0)}S^{(7)}\right)
    +\left(d\Phi_{k}^{(0)}-mP_{B\,k}^{(1)}\right)\wedge S^{(7)}
        \\
    & \\
    &\stackrel{k=0}{=}
      d\left(-\Phi_{k}^{(0)}S^{(7)}\right)\,,
  \end{aligned}
\end{equation}

\noindent
and these troublesome terms, associated to the charge of $C^{(1)}$ will not
contribute wherever $k=0$. On the other hand, since Eq.~(\ref{eq:dphi0})
implies that $P_{B\, k}^{(1)}$ is exact, there will be no contribution
associated to the charge of the KR term, either, which is what we expected in
the massive case, in spite of the fact that one can define conserved charges
for the associated objects. Asymptotically the contributions can be set to
zero by judicious choices of the total derivatives that one can add to the
potentials \cite{Zatti:2024vbz}.

\section{The generalized Komar charge of  $\mathcal{N}=2B,d=10$ supergravity}
\label{sec-n2bd10sugra}

The chiral $\mathcal{N}=2B,d=10$ supergravity theory
\cite{Schwarz:1983qr,Howe:1983sra} contains the same NSNS fields as the $2A$
theory plus 3 RR potentials: $C^{(0)}$, $C^{(2)}$ and $C^{(4)}$. The field
strength of the latter is self-dual: its equation of motion coincides with its
Bianchi identity. For this reason, this theory does not admit an action which
is covariant or makes no use of auxiliary fields as in the PST formalism
\cite{Pasti:1996vs} or in Sen's approach \cite{Sen:2015nph}. Given the
advantages of working with an action instead of the equations of motion, in
Ref.~\cite{Bergshoeff:1995sq} it was proposed to use the action of a different
``non-self-dual'' (NSD) theory\footnote{In Ref.~\cite{Bergshoeff:1995sq} this
  action was called a \textit{pseudoaction}. However, it is more convenient to
  think of it as of the action of a larger ``non-self-dual'' (NSD) theory with
  a subsector that corresponds to the $\mathcal{N}=2B,d=10$ theory.} such that

\begin{enumerate}
\item It has the same field content except for the 4-form RR potential
  $C^{(4)}$, whose 5-form field strength, $G^{(5)}$, would not be constrained
  to be self-dual (hence the name).
\item The equations of motion of the 4-form and the Bianchi identity of its
  5-form field strength are related by the transformation
  $G^{(5)}\rightarrow \star G^{(5)}$. This is required for the restriction
  $G^{(5)}=\star G^{(5)}$ to be consistent.
\item The subsector of the theory that has a self-dual 5-form obeys the original
  equations of motion of the  $\mathcal{N}=2B,d=10$ theory.
\end{enumerate}

This idea is also the basis of the so-called \textit{democratic actions} developed
afterwards and which we will use in the next section.

The action of the NSD theory in string frame\footnote{In this frame the global
  SL$(2,\mathbb{R})$ symmetry of the theory is not manifest at all. It
  involves Weyl rescalings of the string metric, for instance.} has the form
Eq.~(\ref{eq:n2ad10action}) with the RR sector of the Lagrangian given by

\begin{equation}
        \label{eq:LRR2B}
  \begin{aligned}
    \mathbf{L}_{RR}
    & =
      \tfrac{1}{2}G^{(1)}\wedge \star G^{(1)}
      +\tfrac{1}{2}G^{(3)}\wedge \star G^{(3)}
      +\tfrac{1}{4}G^{(5)}\wedge \star G^{(5)}
    \\
    &  \\
    & \hspace{.5cm}
      -\tfrac{1}{2}G^{(5)}\wedge G^{(3)}\wedge B
          +\tfrac{1}{4}G^{(5)}\wedge G^{(1)} \wedge B^{2}
      -\tfrac{1}{2\cdot 3!}G^{(3)}\wedge G^{(1)} \wedge B^{3}\,,
  \end{aligned}
\end{equation}

\noindent
where

\begin{subequations}
  \begin{align}
    G^{(1)}
    & \equiv dC^{(0)}\,,
    \\
    & \nonumber \\
    G^{(3)}
    & \equiv dC^{(2)}-HC^{(0)}\,,
    \\
    & \nonumber \\
    G^{(5)}
    & \equiv dC^{(4)}-H\wedge C^{(2)}\,,
  \end{align}
\end{subequations}

\noindent
are the gauge-invariant RR field strengths of the electric RR potentials
$C^{(0)},C^{(2)}$ and $C^{(4)}$. The gauge transformations that leave them
invariant are

\begin{equation}
  \label{eq:gaugetransformations2B}
\delta_{\Lambda} B = d\Lambda^{(1)}_{B}\,,   
  \hspace{1cm}
\delta_{\Lambda} C^{(2)} = d\Lambda^{(1)}\,,   
  \hspace{1cm}
\delta_{\Lambda} C^{(4)} = d\Lambda^{(3)}-H\wedge \Lambda^{(1)} \,.   
\end{equation}

The field strengths and gauge transformations  fit in the general patterns
Eq.~(\ref{eq:generalRRfieldstrengths}) and
(\ref{eq:generalRRgaugetransformations}), respectively. The equations of
motion will only fit into that pattern after the duality constraint
$G^{(5)}=\star G^{(5)}$ has been imposed, though.

Notice that the kinetic term of the RR 4-form carries an additional factor of
$1/2$ with respect to the other ones. Also, notice that the Chern--Simons term 
in the second line of Eq.~(\ref{eq:LRR2B}) is equivalent to the usual one 

\begin{equation}
-\tfrac{1}{2}dC^{(4)}\wedge dC^{(2)}\wedge B\,,   
\end{equation}

\noindent
(it differs from it by total derivatives), but it is manifestly invariant
under RR gauge transformations. Under the KR gauge transformations, it is
invariant up to a total derivative 

\begin{equation}
  \label{eq:deltaLambdaS2B}
  \begin{aligned}
    \delta_{\Lambda}S
    & =
      -\int_{\mathcal{M}_{10}} d\mathbf{Y}^{(7)}\wedge \delta_{\Lambda}B\,,
    \\
    & \\
    d\mathbf{Y}^{(7)}
    & =
      \tfrac{1}{2} \left.\tilde{G}\right|_{5} \wedge \left.\tilde{G}\right|_{3}
    \\
    & \\
    & =
      \tfrac{1}{2} \left.dS\right|_{4} \wedge \left.dS\right|_{2}\,,
  \end{aligned}
\end{equation}

\noindent
where $\tilde{G}$ and $S$ are the polyforms of Page charges and potentials
defined in Eq.~(\ref{eq:Sdefinitionmassive}) adapted to the odd-rank RR field
strengths we have in this theory and where the subscript indicates the rank at
which is has been truncated, \textit{e.g.}

\begin{equation}
  \left.\tilde{G}\right|_{5}
  =
  G^{(5)} -G^{(3)}\wedge B+\tfrac{1}{2}G^{(1)}\wedge B^{2}\,.
\end{equation}

This means that we can express $\mathbf{Y}^{(7)}$ for instance, as (our
choice, by convention)

\begin{equation}
  \mathbf{Y}^{(7)}
  =
      -\tfrac{1}{2} \left.\tilde{G}\right|_{5} \wedge \left.S\right|_{2}\,.
\end{equation}



Under an arbitrary infinitesimal variation of the fields, the action changes
as

\begin{equation}
  \label{eq:generalvariation2b}
  \begin{aligned}
    \delta S
    & =
    \int_{\mathcal{M}_{10}} \left\{
    \mathbf{E}_{a}\wedge \delta e^{a}
    +\mathbf{E}_{\phi}\delta \phi
    +\mathbf{E}_{B}\wedge \delta B
    \right.
    \\
    & \\
    & \hspace{.5cm}
    \left.
    +\mathbf{E}_{(0)}\wedge \delta C^{(0)}
    +\mathbf{E}_{(2)}\wedge \delta C^{(2)}
    +\mathbf{E}_{(4)}\wedge \delta C^{(4)}
    +d\mathbf{\Theta}(\varphi,\delta \varphi)
    \right\}\,,  
  \end{aligned}
\end{equation}

\noindent
where, by definition,

\begin{subequations}
  \label{eq:eom2B}
  \begin{align}
    \mathbf{E}_{a}
    & =
      e^{-2\phi}\left[\imath_{a}\star (e^{b}\wedge e^{c})\wedge R_{bc}
      -4(\imath_{a}d\phi \star d\phi
      +d\phi\wedge \imath_{a} \star d\phi) \right.
      \nonumber \\
    & \nonumber \\
    & \hspace{.5cm}
      \left.
      +\tfrac{1}{2}(\imath_{a}H \wedge\star H
      +H\wedge \imath_{a} \star H)
      +2e^{2\phi}\mathcal{D}\left(\imath_{a}\star de^{-2\phi}\right)\right]
    \nonumber \\
    & \nonumber \\
    & \hspace{.5cm}
            +\tfrac{1}{2}(\imath_{a}G^{(1)} \wedge\star G^{(1)}
      +G^{(1)}\wedge \imath_{a} \star G^{(1)})
            +\tfrac{1}{2}(\imath_{a}G^{(3)} \wedge\star G^{(3)}
      +G^{(3)}\wedge \imath_{a} \star G^{(3)})
    \nonumber \\
    & \nonumber \\
    & \hspace{.5cm}
            +\tfrac{1}{4}(\imath_{a}G^{(5)} \wedge\star G^{(5)}
      +G^{(5)}\wedge \imath_{a} \star G^{(5)})\,,
      \\
    & \nonumber \\
    \mathbf{E}_{\phi}
    & =
      -4d\star de^{-2\phi} -2\mathbf{L}_{NSNS}\,,
    \\
    & \nonumber \\
    \mathbf{E}_{B}
    & =
      -d(e^{-2\phi}\star H) -\star G^{(3)} \wedge G^{(1)} -G^{(5)+}\wedge G^{(3)}
      -\mathbf{E}_{(2)}C^{(0)} -\mathbf{E}_{(4)}\wedge C^{(2)}\,,
    \\
    & \nonumber \\
    \mathbf{E}_{(0)}
    & =
      -\left(d\star G^{(1)} +H \wedge \star G^{(3)}\right)\,,
    \\
    & \nonumber \\
    \mathbf{E}_{(2)}
    & =
      -\left(d\star G^{(3)} +H \wedge G^{(5)+}\right)\,,
    \\
    & \nonumber \\
    \mathbf{E}_{(4)}
    & =
      -\tfrac{1}{2}\left(d\star G^{(5)} -H \wedge G^{(3)}\right)\,,
  \end{align}
\end{subequations}

\noindent
are the equations of motion of the Vielbein (the Einstein equations), dilaton,
KR and RR fields, with

\begin{equation}
  G^{(5)\pm}
  =
  \tfrac{1}{2}\left(G^{(5)}\pm\star G^{(5)}\right)\,,
\end{equation}

\noindent
and where 

\begin{equation}
  \label{eq:Theta2B}
  \begin{aligned}
    \mathbf{\Theta}(\varphi,\delta \varphi)
    & =
    -e^{-2\phi}\star (e^{a}\wedge e^{b})\wedge \delta \omega_{ab}
    +4\star d e^{-2\phi} \delta\phi
      -2\imath_{a}\star d e^{-2\phi} \wedge \delta e^{a}
    \\
    & \\
    & \hspace{.5cm}
      +\mathbf{Z}^{(7)}\wedge \delta B
      +\mathbf{X}^{(9)}\delta C^{(0)}
      +\mathbf{X}^{(7)}\wedge \delta C^{(2)}
      +\mathbf{X}^{(5)}\wedge \delta C^{(4)}\,,
  \end{aligned}
\end{equation}

\noindent
is the symplectic prepotential.

We have defined 

\begin{subequations}
  \begin{align}
    \mathbf{Z}^{(7)}
    & \equiv
      \frac{\delta S}{\delta H}
      =
      e^{-2\phi}\star H -\mathbf{X}^{(7)}C^{(0)}-\mathbf{X}^{(5)}\wedge C^{(2)}\,,
    \\
    & \nonumber \\
    \mathbf{X}^{(9)}
    & \equiv
      \frac{\delta S}{\delta G^{(1)}}
      =
      \star G^{(1)} -\tfrac{1}{4}G^{(5)}\wedge B^{2}
      +\tfrac{1}{12}G^{(3)}\wedge B^{3}\,, 
    \\
    & \nonumber \\
    \mathbf{X}^{(7)}
    & \equiv
      \frac{\delta S}{\delta G^{(3)}}
      =
      \star G^{(3)} +\tfrac{1}{2}G^{(5)}\wedge B -\tfrac{1}{12} G^{(1)}\wedge B^{3}\,,
    \\
    & \nonumber \\
    \mathbf{X}^{(5)}
    & \equiv
      \frac{\delta S}{\delta G^{(5)}}
      =
      \tfrac{1}{2}
      \left[\star G^{(5)} -G^{(3)}\wedge B +\tfrac{1}{2} G^{(1)}\wedge B^{2}\right]\,.
  \end{align}
\end{subequations}

These forms satisfy the relations

\begin{subequations}
  \label{eq:XYZrelations2B}
  \begin{align}
    d\left(\mathbf{Z}^{(7)}+\mathbf{Y}^{(7)}\right)
    & =
      -\mathbf{E}_{B}\,,
    \\
    & \nonumber \\
    d\mathbf{X}^{(9)} +H\wedge \mathbf{X}^{(7)}
    & =
      -\mathbf{E}_{(0)}\,,
    \\
    & \nonumber \\
    d\mathbf{X}^{(7)} +H\wedge \mathbf{X}^{(5)}
    & =
      -\mathbf{E}_{(2)}\,,
    \\
    & \nonumber \\
    d\mathbf{X}^{(5)}
    & =
      -\mathbf{E}_{(4)}\,.
  \end{align}
\end{subequations}

In order to recover the equations of motion or the on-shell conserved charges
of the original $\mathcal{N}=2B,d=10$ supergravity theory we just have to
impose the self-duality constraint

\begin{equation}
G^{(5)}=\star G^{(5)}\,,  
\end{equation}

\noindent
on them. As it is well known, imposing this constraint in the action removes
the kinetic term of $C^{(4)}$ and leads to an action from which one cannot
derive the equations of motion. Evaluating the NSD action on-shell is, then,
problematic.

\subsection{Symmetries and conserved charges}
\label{sec-n2bd10sugrasymmetriesandcharges}

\subsubsection{Local Lorentz transformations}
\label{sec-n2bd10LL}

Since the only Lorentz tensor fields of this theory belong to the NSNS sector
which is common to the $\mathcal{N}=2A$ and $\mathcal{N}=2B$ theories, the
same results found and comments made in Section~\ref{sec-n2ad10LL} apply here
and we will not repeat them.

\subsubsection{NSNS and RR gauge transformations}
\label{sec-n2bd10gauge}

We follow the same steps as in the previous cases, computing first

\begin{equation}
  \begin{aligned}
    \delta_{\Lambda} S
    & =
    \int_{\mathcal{M}_{10}} \left\{
    -d\mathbf{E}_{B}\wedge  \Lambda^{(1)}_{B}
    -\left(d\mathbf{E}_{(2)}+\mathbf{E}_{(4)}\wedge H\right)\wedge \Lambda^{(1)}
      -d\mathbf{E}_{(4)}\wedge \Lambda^{(3)}
      \right.
    \\
    & \\
    & \hspace{.5cm}
      \left.
      +d\left[\mathbf{\Theta}(\varphi,\delta_{\Lambda} \varphi)
      +\mathbf{E}_{B}\wedge \Lambda^{(1)}_{B}
      +\mathbf{E}_{(2)}\wedge \Lambda^{(1)}
      +\mathbf{E}_{(4)}\wedge \Lambda^{(3)}
\right]
      \right\}\,.    
  \end{aligned}
\end{equation}

The first line vanishes identically because each of the terms proportional to the
gauge transformation parameters vanishes identically due to the Noether identities.

On the other hand, from Eq.~(\ref{eq:deltaLambdaS2B}) we have

\begin{equation}
    \delta_{\Lambda}S
     =
      -\int_{\mathcal{M}_{10}} d\mathbf{Y}^{(7)}\wedge d\Lambda^{(1)}_{B}\,,
\end{equation}

\noindent
and we arrive at the off-shell conserved Noether current

\begin{equation}
  \mathbf{J}[\Lambda]
  =
  \mathbf{\Theta}(\varphi,\delta_{\Lambda} \varphi)
      +\mathbf{E}_{B}\wedge \Lambda^{(1)}_{B}
      +\mathbf{E}_{(2)}\wedge \Lambda^{(1)}
      +\mathbf{E}_{(4)}\wedge \Lambda^{(3)}
      +\mathbf{Y}^{(7)}\wedge d\Lambda^{(1)}_{B}\,,
\end{equation}

\noindent
which, after some work, can be expressed as a total derivative

\begin{subequations}
  \begin{align}
    \mathbf{J}[\Lambda]
    & =
      d\mathbf{Q}[\Lambda]\,,
    \\
    & \nonumber \\
    \mathbf{Q}[\Lambda]
    & =
      -\left(\mathbf{Z}^{(7)}+\mathbf{Y}^{(7)}\right)\wedge \Lambda^{(1)}_{B}
      -\left(\mathbf{X}^{(7)}+B\wedge  \mathbf{X}^{(5)}\right)\wedge \Lambda^{(1)}
      \nonumber \\
    & \nonumber \\
    & \hspace{.5cm}
      -\mathbf{X}^{(5)}\wedge \left(\Lambda^{(3)}-B\wedge \Lambda^{(1)}\right)\,.
  \end{align}
\end{subequations}

Using the magnetic RR field strengths\footnote{These definitions, which
  include additional terms proportional to $G^{(5)-}$ that will vanish when
  the self-duality constraint is imposed, follow directly from on-shell
  dualization.}

\begin{subequations}
  \begin{align}
    G^{(7)} & \equiv
              -\star G^{(3)}
              =
              dC^{(6)}-H\wedge C^{(4)} -B\wedge G^{(5)-}\,,
    \\
    & \nonumber \\
    G^{(9)} & \equiv
              \star G^{(1)}
              =
              dC^{(8)}-H\wedge C^{(6)} -\tfrac{1}{2}B^{2}\wedge G^{(5)-}\,,
  \end{align}
\end{subequations}

\noindent
we find 

\begin{subequations}
  \begin{align}
    \mathbf{Z}^{(7)}+\mathbf{Y}^{(7)}
    & =
      e^{-2\phi}\star H -\tfrac{1}{2}\star G\wedge C
      -\tfrac{1}{2}G^{(5)-}\wedge BC^{(0)}
      \nonumber     \\
    & \nonumber \\
    & \hspace{-2cm}
     +\tfrac{1}{2}d\left\{C^{(0)}C^{(6)} -C^{(2)}\wedge \left(C^{(4)} -B\wedge
      C^{(2)} +B^{2}C^{(0)}\right)
      +\tfrac{1}{3}B^{3}C^{(0)}C^{(0)} \right\} \,,
    \\
    & \nonumber \\
    \mathbf{X}^{(7)}+B\wedge  \mathbf{X}^{(5)}
    & =
      -G^{(7)} +G^{(5)+}\wedge B -\tfrac{1}{2}G^{(3)}\wedge B^{2}
      +\tfrac{1}{3!} G^{(1)}\wedge      B^{3}\,,
      \\
    & \nonumber \\
    \mathbf{X}^{(5)}
    & =
      \tfrac{1}{2}
      \left[\star G^{(5)} -G^{(3)}\wedge B +\tfrac{1}{2} G^{(1)}\wedge B^{2}\right]
  \end{align}
\end{subequations}

\noindent
which, up to terms proportional to $G^{(5)-}$ and total derivatives,
correspond to the generic RR Page charges and agree with the form of the KR
Page charge of the massless $\mathcal{N}=2A$ theory, discussed in
Section~\ref{sec-n2bd10gauge}, up to the obvious differences in the
composition of the polyforms $G$ and $C$.

\subsubsection{General coordinate transformations}
\label{sec-n2bd10gct}

We just quote the results of the intermediate steps in our calculations since
the definitions of the parameters of the compensating gauge transformations,
Bianchi identities, momentum maps equations and $\delta_{\xi}$ transformations
of all the NSNS and electric RR fields ($C^{(0)},C^{(2)},C^{(4)}$) are
qualitatively the same as in the previous cases.

After the Noether identities have been used, the Noether current associated to
GCTs consists of the following terms:

\begin{equation}
  \begin{aligned}
    \mathbf{J}[\xi]
    & =
    \mathbf{\Theta}(\varphi,\delta_{\xi} \varphi)
    +\xi^{a}\left(\mathbf{E}_{a}+\imath_{a}\mathbf{L}\right)
    -\Lambda^{(1)}_{B}\wedge \left(\mathbf{E}_{B}
      +\mathbf{E}_{(2)}C^{(0)}+\mathbf{E}_{(4)}\wedge C^{(0)}\right)
    \\
    & \\
    & \hspace{.5cm}
    -P_{\xi}^{(1)}\wedge \mathbf{E}_{(2)}
    -P_{\xi}^{(3)}\wedge \mathbf{E}_{(4)}\,.
  \end{aligned}
\end{equation}

\noindent
Massaging them, we find the Noether--Wald charge

\begin{equation}
  \begin{aligned}
    \mathbf{Q}_{NW}[\xi]
    & =
    e^{-2\phi} \star (e^{a}\wedge e^{b})P_{\xi\, ab}
    +2\imath_{\xi}\star de^{-2\phi} -P_{B\, \xi}^{(1)}\wedge (e^{-2\phi}\star
      H)
    \\
    & \\
    & \hspace{.5cm}
    -P_{\xi}^{(1)}\wedge \mathbf{X}^{(7)}
    -P_{\xi}^{(2)}\wedge \mathbf{X}^{(5)}\,.
  \end{aligned}
\end{equation}

In order to compute $\imath_{k}\mathbf{L}_{on-shell}$ we cannot use the
generic equation for magnetic RR momentum maps because it is based on the generic
form of the equations of motion (\ref{eq:genericRReoms}) which is not obeyed
in the NSD theory. We, instead, find the following magnetic momentum map
equations

\begin{subequations}
  \begin{align}
  \imath_{k}\star G^{(5)} -P_{B\, k}^{(1)}  \wedge G^{(3)}
    +d\tilde{P}_{k}^{(3)} -H\wedge P_{k}^{(1)}
    & =
      0\,,
    \\
    & \nonumber \\
  \imath_{k}\star G^{(3)} +P_{B\, k}^{(1)}  \wedge G^{(5)+}
    +d\tilde{P}_{k}^{(5)} +H\wedge \tfrac{1}{2}\left(P_{k}^{(3)}+\tilde{P}_{k}^{(3)} \right)
    & =
      0\,,
    \\
    & \nonumber \\
      \imath_{k}\star G^{(1)} +P_{B\, k}^{(1)}  \wedge \star G^{(3)}
    +d\tilde{P}_{k}^{(7)} +H\wedge P_{k}^{(5)}
    & =
      0\,.
  \end{align}
\end{subequations}

When the self-duality constraint is satisfied,
$P_{k}^{(3)}=\tilde{P}_{k}^{(3)}$, and these equations take the standard form.

We obtain

\begin{subequations}
  \begin{align}
d\mathbf{w}_{k}
    & \equiv     \imath_{k}\mathbf{L}_{on-shell}
    +d \mathbf{Y}^{(7)} \wedge \Lambda^{(1)}_{B\, k}\,,
    \\
    & \nonumber \\
    \mathbf{w}_{k}
    & =
      2\imath_{k}\star d e^{-2\phi}
      -\tfrac{1}{2}P_{k}^{(1)}\wedge
      \left(\star G^{(3)}+G^{(5)}\wedge B -\tfrac{1}{3!} G^{(1)}\wedge B^{3}
      \right)
      \nonumber \\
    & \nonumber \\
    & \hspace{.5cm}
      -\tfrac{1}{4}P_{k}^{(3)}\wedge
      \left(\star G^{(5)} -2G^{(3)}\wedge B +G^{(1)}\wedge B^{2}\right)
      \nonumber \\
    & \nonumber \\
    & \hspace{.5cm}
      +\tfrac{1}{4}\tilde{P}_{k}^{(3)}\wedge G^{(5)}
      -\tfrac{1}{2}P_{k}^{(5)}\wedge G^{(3)}
      +\tfrac{1}{2}P_{k}^{(7)}\wedge G^{(1)}\,,
  \end{align}
\end{subequations}

\noindent
so that the on-shell conserved generalized Komar form of the NSD theory is

\begin{equation}
  \label{eq:Komarcharge2BNSD}
  \begin{aligned}
    \mathbf{K}[k]
    & \equiv
      -\mathbf{Q}_{NW}[k]+\mathbf{w}_{k}
    \\
    & \\
    & =
    -e^{-2\phi} \star (e^{a}\wedge e^{b})P_{k\, ab}
    +P_{B\, k}^{(1)}\wedge (e^{-2\phi}\star H)
    +\tfrac{1}{2}P_{k}^{(1)}\wedge \star G^{(3)}
    \\
    & \\
    & \hspace{.5cm}
    +\tfrac{1}{4}P_{k}^{(3)}\wedge \star G^{(5)}
    +\tfrac{1}{4}\tilde{P}_{k}^{(3)}\wedge G^{(5)}
    -\tfrac{1}{2}P_{k}^{(5)}\wedge G^{(3)}
    +\tfrac{1}{2}P_{k}^{(7)}\wedge G^{(1)}\,,
  \end{aligned}
\end{equation}

\noindent
and, when the self-duality constraint $G^{(5)}=\star G^{(5)}$ is used, we get
the on-shell conserved generalized Komar charge of the original
$\mathcal{N}=2B,d=10$ supergravity theory:

\begin{equation}
  \label{eq:Komarcharge2B}
  \begin{aligned}
    \mathbf{K}[k]
    & =
    -e^{-2\phi} \star (e^{a}\wedge e^{b})P_{k\, ab}
    +P_{B\, k}^{(1)}\wedge (e^{-2\phi}\star H)
    +\tfrac{1}{2}P_{k}\wedge \star G\,.
  \end{aligned}
\end{equation}

Observe that it has exactly the same form as that of the massless and massive
$\mathcal{N}=2A$ theories Eq.~(\ref{eq:Komarcharge2A}), but in this case the
polyforms only contain odd-rank RR field strengths and odd-rank RR momentum
maps.

Following the same steps as in the $\mathcal{N}=2A$ case, we can rewrite it
completely in terms of potentials that are closed when $k=0$ and on-shell
closed Page charges. The result has the structure of
Eq.~(\ref{eq:Komarcharge2APage}) with the differences in the polyforms
mentioned before.

\section{The generalized Komar charge from the democratic formulation}
\label{sec-democratic}

A very convenient formulation of both $\mathcal{N}=2,d=10$ theories in which
all RR fields (including their magnetic duals), but not the NSNS
ones,\footnote{In this RR-democratic formulation, the SL$(2,\mathbb{R})$
  duality group of the $\mathcal{N}=2B,d=10$ theory that involves RR and NSNS
  fields is not manifest. A fully democratic and manifestly
  SL$(2,\mathbb{R})$-covariant formulation of the $\mathcal{N}=2B,d=10$ theory
  has recently been given in Ref.~\cite{Fernandez-Melgarejo:2023kwk}.} are
treated on equal footing as independent variables, was proposed in
Ref.~\cite{Bergshoeff:2001pv}. If we use the polyforms
$C \equiv \sum_{n=0}^{9}C^{(n)}$ and $G \equiv \sum_{n=0}^{10}G^{(n)}$ for the
RR potentials and field strengths, we can write the action in the form of
Eq.~(\ref{eq:n2ad10action}) with a different RR sector given by

\begin{equation}
  \label{eq:democraticaction}
    \mathbf{L}_{RR}=\tfrac{1}{4}G\wedge \star G\,.
\end{equation}

The KR field strength and the KR gauge transformations have the same
expression as in the $A,B$ theories and we reproduce them here for the sake of
convenience, as we will do with several identities that will be used in this
section:

\begin{equation}
    H\equiv dB\,,
    \hspace{1cm}
    \delta_{\Lambda} B = d\Lambda^{(1)}_{B}\,, 
\end{equation}

The RR field strengths follow the general definition in
Eq.~(\ref{eq:generalRRfieldstrengths}) and satisfy the Bianchi identities
Eq.~(\ref{eq:generalRRBianchiidentities}) and the gauge transformations of the
RR potentials also have the general form given in
Eq.~(\ref{eq:generalRRgaugetransformations})

\begin{equation}
  G = d_{H^{-}}C +me^{B}\,,
  \hspace{1cm}
  d_{H^{-}}G = 0\,,
  \hspace{1cm}
  \delta_{\Lambda}C
  =
  d_{H^{-}}\Lambda -m e^{B}\wedge \Lambda_{B}^{(1)}\,.
\end{equation}

Here we are using the operators

\begin{equation}
    d_{H^{\pm}}
    \equiv d\pm H\wedge\,,
    \hspace{1cm}
    d_{H^{\pm}}^{2} = 0\,.
\end{equation}

The action of the democratic theory is exactly gauge invariant because it does
not have any Chern--Simons terms.

Under an arbitrary infinitesimal variation of the fields, the democratic
action changes as

\begin{equation}
  \label{eq:generalvariationdemo}
  \delta S
  =
  \int_{\mathcal{M}_{10}} \left\{
    \mathbf{E}_{a}\wedge \delta e^{a}
    +\mathbf{E}_{\phi}\delta \phi
    +\mathbf{E}_{B}\wedge \delta B
    +\mathbf{E}\wedge \delta C
    +d\mathbf{\Theta}(\varphi,\delta \varphi)
  \right\}\,,  
\end{equation}

\noindent
where we are using the polyform of RR equations of motion $\mathbf{E}$ defined
in the obvious way.

The equations of motion take the explicit form

\begin{subequations}
  \label{eq:eom2demo}
  \begin{align}
    \mathbf{E}_{a}
    & =
      e^{-2\phi}\left[\imath_{a}\star (e^{b}\wedge e^{c})\wedge R_{bc}
      -4(\imath_{a}d\phi \star d\phi
      +d\phi\wedge \imath_{a} \star d\phi) \right.
      \nonumber \\
    & \nonumber \\
    & \hspace{.5cm}
      \left.
      +\tfrac{1}{2}(\imath_{a}H \wedge\star H
      +H\wedge \imath_{a} \star H)
      +2e^{2\phi}\mathcal{D}\left(\imath_{a}\star de^{-2\phi}\right)\right]
    \nonumber \\
    & \nonumber \\
    & \hspace{.5cm}
            +\tfrac{1}{4}\sum_{n=0}^{10}(\imath_{a}G^{(n)} \wedge\star G^{(n)}
      -(-1)^{n}G^{(n)}\wedge \imath_{a} \star G^{(n)})\,,
      \\
    & \nonumber \\
    \mathbf{E}_{\phi}
    & =
      -4d\star de^{-2\phi} -2\mathbf{L}_{NSNS}\,,
    \\
    & \nonumber \\
    \mathbf{E}_{B}
    & =
      -d(e^{-2\phi}\star H) +\tfrac{1}{2}G\wedge \star G -\mathbf{E}\wedge C\,,
    \\
    & \nonumber \\
    \mathbf{E}
    & =
      -\tfrac{1}{2} d_{H^{+}}\star G\,,
  \end{align}
\end{subequations}

\noindent
while the presymplectic potential is given by

\begin{equation}
  \label{eq:Thetademo}
  \begin{aligned}
    \mathbf{\Theta}(\varphi,\delta \varphi)
    & =
    -e^{-2\phi}\star (e^{a}\wedge e^{b})\wedge \delta \omega_{ab}
    +4\star d e^{-2\phi} \delta\phi
      -2\imath_{a}\star d e^{-2\phi} \wedge \delta e^{a}
    \\
    & \\
    & \hspace{.5cm}
      +\mathbf{Z}^{(7)}\wedge \delta B
      +\tfrac{1}{2}\star G\wedge \delta C\,,
  \end{aligned}
\end{equation}

\noindent
where we have defined the 7-form

\begin{equation}
  \label{eq:Z7demodef}
  \mathbf{Z}^{(7)}
  \equiv
        \frac{\delta S}{\delta H}
      =
      e^{-2\phi}\star H -\tfrac{1}{2} C\wedge \star G\,,
\end{equation}

\noindent
which satisfies

\begin{equation}
    \label{eq:Z7demoproperty}
    d\mathbf{Z}^{(7)} -\tfrac{1}{2}m e^{B}\wedge \star G
    =
    d\left(\mathbf{Z}^{(7)} -\tfrac{1}{2}m \tilde{S}^{(7)}\right)
    =
    -\mathbf{E}_{B}\,.
\end{equation}

The equations of motion of the original $\mathcal{N}=2A,B$ theories are
recovered in two steps

\begin{enumerate}
\item Using first the duality constraints Eq.~(\ref{eq:dualityconstraints})

\begin{equation}
G^{(10-n)} = (-1)^{[n/2]} \star G^{(n)}\,.
\end{equation}
  
\item Keeping only the even (odd) $n$ RR field strengths for the $2A$ ($2B$)
  theory.

\end{enumerate}

In the same spirit of the NSD theory used in Section~\ref{sec-n2bd10sugra} to
derive the equations of motion and the generalized Komar charge of the
$\mathcal{N}=2B,d=10$ supergravity theory, the ``democratic'' theory
Eq.~(\ref{eq:democraticaction}) should be understood as a different theory
larger than the $\mathcal{N}=2A,B,d=10$ ones. These correspond to the subsectors
of the democratic theory satisfying the duality constraints
Eq.~(\ref{eq:dualityconstraints}). All the properties of the original theories
that only depend on the equations of motion can be derived from the democratic
formulation by applying the duality constraints.

In the calculations that lead to the different conserved charges  of the
democratic theory and which follow the procedure established in the previous
sections without the complications associated to the Chern--Simons term of the
action, we will make use of

\begin{enumerate}

\item The ``magnetic'' RR Page charges and associated potentials 

\begin{equation}
  \tilde{G}
  \equiv
  G\wedge e^{-B}
  =
  d(C\wedge e^{-B}) +m
  \equiv
  dS +m\,.
\end{equation}

\item The ``electric'' RR Page charges and associated potentials

\begin{equation}
  \widetilde{\star G}
  \equiv
  \star G\wedge e^{B}
  \equiv
  d\tilde{S}+\tilde{m}\,,
\end{equation}

\noindent
where $+\tilde{m}$ is a scalar, not necessarily constant, since in the
democratic theory $G^{(10)}$ is unrelated to $G^{(0)}=m$.

\item The ``electric'' KR Page charge

\begin{equation}
\tilde{H}    
\equiv
e^{-2\phi}\star H -\tfrac{1}{2}C\wedge \star G -\tfrac{1}{2}m\tilde{S}^{(7)}
=
\mathbf{Z}^{(7)} -\tfrac{1}{2}m\tilde{S}^{(7)}\,,
\end{equation}

\noindent
where $\mathbf{Z}^{(7)}$ has been defined in Eq.~(\ref{eq:Z7demodef}) and we
should remember that it satisfies Eq.~(\ref{eq:Z7demoproperty}).

\item The ``electric'' KR and RR momentum maps $P^{(1)}_{B\, k}$ and
  $P_{k}^{(n)}$, which are defined by the momentum map equations
  (\ref{eq:KRmomentummapeq})

\begin{equation}
  \imath_{k}H +dP^{(1)}_{B\, k} = 0\,,
\end{equation}

\noindent
and (\ref{eq:RRgeneralmomentummapeq})

\begin{equation}
  \imath_{k}G -P^{(1)}_{B\, k}\wedge G +d_{H^{-}}P_{k}=0\,,
  \hspace{1cm}
  P_{k}\equiv \sum_{n=0}^{8}P_{k}^{(n)}\,.
\end{equation}

\item  The ``magnetic'' RR momentum maps $\tilde{P}_{k}^{(n)}$ which,
combined into the polyform $\tilde{P}_{k}=\sum_{n=0}^{8}\tilde{P}_{k}^{(n)}$
satisfy Eq.~(\ref{eq:RRmagneticmomentummapequations})

\begin{equation}
  \imath_{k}\star G +P^{(1)}_{B\, k}\wedge \star G +d_{H^{+}}\tilde{P}_{k}=0\,.    
\end{equation}

\item The potentials

\begin{subequations}
  \begin{align}
    \Phi_{B\, k}^{(1)}
    & \equiv  P_{B\, k}^{(1)}\,,
    \\
    & \nonumber \\
    \Phi_{k}
    & \equiv
      \left(P_{k}+P_{B\, k}^{(1)}\wedge C\right)\wedge e^{-B}\,,
  \end{align}
\end{subequations}

\noindent
which satisfy

\begin{subequations}
  \begin{align}
    d\Phi_{B\, k}^{(1)}
    & \stackrel{k=0}{=}0\,,
    \\
    & \nonumber \\
    d\Phi_{k}
    & \stackrel{k=0}{=}
      m \Phi_{B\, k}^{(1)}\,,
  \end{align}
\end{subequations}

\noindent
so all of them are closed on Killing horizons except for $\Phi_{k}^{(0)}$,
which makes $\Phi_{B\, k}^{(1)}$ exact.

\end{enumerate}

The electric and magnetic momentum maps, Page charges and associated
potentials of the democratic theory are independent but, when the duality
constraints Eq.~(\ref{eq:dualityconstraints}) hold, they are related by
Eq.~(\ref{eq:dualityconstraintsRRmomentummaps}),
(\ref{eq:dualityconstraintsPagecharges}) and
(\ref{eq:dualityconstraintsPagepotentials}):

\begin{subequations}
  \begin{align}
    \tilde{P}_{k}^{(n)} & =  (-1)^{[(8-n)/2]}  P_{k}^{(n)}\,,
    \\
                        & \nonumber \\
    \tilde{G}^{(10-n)} & = (-1)^{[n/2]} \widetilde{\star G^{(n)}}\,,
    \\
                        & \nonumber \\
    S^{(n)} & = (-1)^{[(9-n)/2]} \tilde{S}^{(n)}\,.
  \end{align}
\end{subequations}

\subsection{Symmetries and conserved charges}
\label{sec-n2d10demosugrasymmetriesandcharges}

\subsubsection{Local Lorentz transformations}
\label{sec-n2d10demoLL}

The situation of the democratic theory with respect to local Lorentz
transformations is exactly the same as in the massless $\mathcal{N}=2A,d=10$
case studied in Section~\ref{sec-n2ad10LL}, to which we refer the reader.

\subsubsection{NSNS and RR gauge transformations}
\label{sec-n21d0demogauge}

The variation of the democratic action under arbitrary gauge transformations
is, after integration by parts and the use of Noether identities

\begin{equation}
  \begin{aligned}
    \delta_{\Lambda} S
    & =
    \int_{\mathcal{M}_{10}} d\left[\mathbf{\Theta}(\varphi,\delta_{\Lambda}\varphi)
      +\mathbf{E}_{B}\wedge  \Lambda^{(1)}_{B}
      +(-1)^{C}\mathbf{E}\wedge \Lambda \right]\,,
  \end{aligned}
\end{equation}

\noindent
and the conserved current can be expressed as a the total derivative of a charge

\begin{equation}
  \begin{aligned}
    \mathbf{J}[\Lambda]
    & =
      d\mathbf{Q}[\Lambda]\,,
    \\
      & \nonumber \\
    \mathbf{Q}[\Lambda]
      & =
           -\mathbf{Z}^{(7)}\wedge \Lambda^{(1)}_{B}
        +(-1)^{C+1}\tfrac{1}{2}\star G\wedge \Lambda
    \\
      & \nonumber \\
    & =
            \Lambda^{(1)}_{B} \wedge \left(\tilde{H} +\tfrac{1}{2}m\tilde{S}^{(7)} \right)
        +\tfrac{1}{2}\tilde{\Lambda}\wedge \widetilde{\star G}\,,
  \end{aligned}
\end{equation}

\noindent
which we have rewritten in the second line in terms of the Page charges
defined above and the gauge parameters $\tilde{\Lambda}=\Lambda\wedge
e^{-B}$. In the above and following expressions $C$ is the degree of the RR
potential.

These charges are conserved for Killing parameters that leave invariant all
the fields. These Killing parameters must satisfy

\begin{subequations}
  \begin{align}
    d\Lambda_{B}^{(1)}
    & =
      0\,,
    \\
    & \nonumber \\
    d\tilde{\Lambda}^{(n)}
    & =
      0\,,
      \,\,\,\, n\neq 0\,,
    \\
    & \nonumber \\
    d\tilde{\Lambda}^{(0)}
    & =
      m \Lambda_{B}^{(1)}\,.
  \end{align}
\end{subequations}

This suggests the definitions:

\begin{equation}
    \mathbf{Q}[\Lambda_{B}^{(1)}]
    \equiv
      \Lambda_{B}^{(1)}\wedge \tilde{H}\,,
\end{equation}

\noindent
for the electric KR charge, which is conserved on-shell and non-trivial when
$\Lambda_{B}^{(1)}$ is closed but not exact,

\begin{equation}
    \mathbf{Q}[\tilde{\Lambda}^{(n)}]
    \equiv
      \tilde{\Lambda}^{(n)}\wedge \left.\widetilde{\star G}\right|_{8-n}\,,  
\end{equation}

\noindent
for the electric RR charges with $n=1,\cdots,8$. These are conserved on-shell
and non-trivial when the $n$-forms $\tilde{\Lambda}^{(n)}$ are closed but not
exact.

Finally, we may also define

\begin{equation}
\mathbf{Q}[\tilde{\Lambda}^{(0)}]
  \equiv
  \tilde{\Lambda}^{(0)}
  \left.\widetilde{\star  G}\right|_{8}+m\Lambda_{B}^{(1)}\wedge \tilde{S}^{(7)}\,,
\end{equation}

\noindent
which is conserved on-shell when
$m\Lambda_{B}^{(1)}=d\tilde{\Lambda}^{(0)}$. When this happens, though, the KR
charge is trivial and

\begin{equation}
\mathbf{Q}[\tilde{\Lambda}^{(0)}]
  =
  \tilde{\Lambda}^{(0)}
  \left.\widetilde{\star  G}\right|_{8}
  +d\tilde{\Lambda}^{(0)}\wedge \tilde{S}^{(7)}
  =
  d\left(\tilde{\Lambda}^{(0)} \tilde{S}^{(7)}\right)\,,
\end{equation}

\noindent
is trivial as well. This combination will naturally occur in the Komar charge,
as we have seen in Section~\ref{sec-n2amd10gct}. We can define, however, the
on-shell conserved charge

\begin{equation}
\mathbf{Q}[\tilde{\Lambda}^{(0)}=1]
  \equiv
\left.\widetilde{\star  G}\right|_{8}\,,
\end{equation}

\noindent
which is compatible with a closed but not exact $\Lambda_{B}^{(1)}$ because it
does not depend on it and, as a result, we see that we can define independent
charges associated to all the gauge parameters.

We can also define magnetic RR and KR charges using the $\tilde{G}^{(n)}$s,
and $H$ in the obvious way.

\subsubsection{General coordinate transformations}
\label{sec-n2d10demogct}

The transformations $\delta_{\xi}$ of all the fields have been determined in
the previous sections and we can proceed to calculate the variation of the
action under them. After integration by parts, we find

\begin{equation}
  \begin{aligned}
    \delta_{\xi}S
    & =
      \int_{\mathcal{M}_{10}} \left\{
      d\left[\mathbf{\Theta}(\varphi,\delta \varphi)
      +\xi^{a}\mathbf{E}_{a}
      -P_{B\, \xi}^{(1)}\wedge \left(\mathbf{E}_{B}+\mathbf{E}\wedge C\right)
      -(-1)^{C}P_{\xi}\wedge \mathbf{E}\right]
      \right.
    \\
    & \\
    & \hspace{.5cm}
      +\xi^{a}\left[-\mathcal{D}\mathbf{E}_{a}
      -\mathbf{E}_{\phi}\imath_{a}d\phi
      - \left(\mathbf{E}_{B}+\mathbf{E}\wedge C\right)\wedge \imath_{a}H
      -\mathbf{E}\wedge \imath_{a}G
      \right]
    \\
    & \\
    & \hspace{.5cm}
      +P_{B\, \xi}^{(1)}\wedge
      \left[-d\left(\mathbf{E}_{B}+\mathbf{E}\wedge C\right)
      +(-1)^{C}\mathbf{E}\wedge G\right]
    \\
    & \\
    & \hspace{.5cm}
      \left.
      -P_{\xi}\wedge d_{H^{+}}\mathbf{E}
      \right\}\,.
  \end{aligned}
\end{equation}

The terms in the last three lines vanish independently because of the Noether
identities and, therefore, we arrive at the off-shell conserved Noether
current

\begin{equation}
  \mathbf{J}[\xi]
  =
  \mathbf{\Theta}(\varphi,\delta \varphi)
      +\xi^{a}\left(\mathbf{E}_{a}+\imath_{a}\mathbf{L}\right)
      -P_{B\, \xi}^{(1)}\wedge \left(\mathbf{E}_{B}+\mathbf{E}\wedge C\right)
      -(-1)^{C}P_{\xi}\wedge \mathbf{E}\,,
\end{equation}

\noindent
which, with the help of the definition of the electric RR momentum maps, can
be written as the total derivative of the Noether--Wald 8-form

\begin{subequations}
  \begin{align}
  \mathbf{J}[\xi]
  & =
    d \mathbf{Q}_{NW}[\xi]\,,
    \\
  & \nonumber \\
    \mathbf{Q}_{NW}[\xi]
  & =
      e^{-2\phi} \star (e^{a}\wedge e^{b})P_{\xi\, ab}
  +2\imath_{\xi}\star de^{-2\phi} -P_{B\, \xi}^{(1)}\wedge (e^{-2\phi}\star H)
-\tfrac{1}{2}P_{\xi}\wedge \star G\,.
  \end{align}
\end{subequations}

On-shell and for a Killing vector $\xi=k$

\begin{equation}
  d \mathbf{Q}_{NW}[k]
  =
  \imath_{k}\mathbf{L}_{on-shell}\,.
\end{equation}

The term in the right-hand side of this equation can be rewritten using the
dilaton equation of motion and the magnetic momentum map equations as another
total derivative

\begin{subequations}
  \begin{align}
    \imath_{k}\mathbf{L}_{on-shell}
    & =
      d\mathbf{w}_{k}\,,
    \\
    & \nonumber \\
    \mathbf{w}_{k}
    & =
          2\imath_{k}\star d e^{-2\phi}
      -\tfrac{1}{4}P_{k}\wedge \star G
      +\tfrac{1}{4}(-1)^{C} \tilde{P}_{k}\wedge G\,,
  \end{align}
\end{subequations}

\noindent
and we arrive at the on-shell conserved generalized Komar charge of the
democratic theory

\begin{equation}
  \label{eq:Komarchargedemocratic}
  \begin{aligned}
    \mathbf{K}[k]
    & \equiv
    -\mathbf{Q}_{NW}[k]+\mathbf{w}_{k}
    =
    -e^{-2\phi} \star (e^{a}\wedge e^{b})P_{k\, ab}
      +P_{B\, k}^{(1)}\wedge (e^{-2\phi}\star H)
    \\
    & \\
    & \hspace{.5cm}
    +\tfrac{1}{4}P_{k}\wedge \star G
    +\tfrac{1}{4}(-1)^{C} \tilde{P}_{k}\wedge G\,.
  \end{aligned}
\end{equation}

Using the duality constraints  of the RR field strengths and electric and
magnetic momentum maps we can eliminate the magnetic ones, obtaining

\begin{equation}
  \label{eq:Komarchargedemocratic2}
  \mathbf{K}[k]
  \equiv
  -\mathbf{Q}_{NW}[k]+\mathbf{w}_{k}
  =
  -e^{-2\phi} \star (e^{a}\wedge e^{b})P_{k\, ab}
  +P_{B\, k}^{(1)}\wedge (e^{-2\phi}\star H)
  +\tfrac{1}{2}P_{k}\wedge \star G\,,
\end{equation}

\noindent
which has the general form we obtained in the massless and massive $2A$ and in
the $2B$ theories. Now we just have to remove the odd- or even-rank RR field
strengths to recover the results obtained before.

We can rewrite the generalized Komar charge in terms of potentials and Page
charges as in the previous cases. It has the same form as in the massive
$\mathcal{N}=2A$ theory, Eq.~(\ref{eq:Komarcharge2AmassivePage}), although the
polyforms include more fields.

\begin{equation}
  \label{eq:KomarchargedemocraticPage}
    \mathbf{K}[k]
    =
    -e^{-2\phi} \star (e^{a}\wedge e^{b})P_{k\, ab}
      +P_{B\, k}^{(1)}\wedge \tilde{H}
      +\tfrac{1}{2}\Phi_{k} \wedge  \widetilde{\star G}
  +\tfrac{1}{2}mP_{B\, k}^{(1)}\wedge \tilde{S}^{(7)}\,
\end{equation}

\noindent
and the same observation involving the non-standard term in Killing horizons
applies to it.

\section{Unbroken supersymmetry and the generalized Komar charge}
\label{sec-unbroken}

In this section we are going to show that, for a supersymmetric Killing vector
(a Killing vector constructed as a bilinear of a Killing spinor), and for
specific choices of the momentum maps which are singled out by supersymmetry
(they are bilinears of the Killing spinor as well) the generalized Komar charges
that we have computed in the previous sections vanish identically.

Let us start by reviewing the Killing spinor equations of these theories. It
is most convenient to use the results of Ref.~\cite{Bergshoeff:2001pv} in
which the supersymmetry transformation rules of the two $\mathcal{N}=2A,B$
theories are be formulated in a democratic way .\footnote{These cannot be the
  supersymmetry transformation rules of the democratic theory because, if that
  was the case, there would be another $\mathcal{N}=2,d=10$ supergravity apart
  from the known ones. Thus, the supersymmetry transformation rules presented
  in Ref.~\cite{Bergshoeff:2001pv} should be considered as a unified way to
  write those of both theories.} Here we are only interested in those of the
fermionic fields of the theory, gravitinos $\psi_{\mu}$ and dilatinos
$\lambda$, and only for purely bosonic field configurations.

We are going to use pairs of Majorana--Weyl spinors labeled by an index that
we will not always show explicitly. The supersymmetry transformation parameter
$\epsilon$ of Ref.~\cite{Bergshoeff:2001pv} will be decomposed in the sum of
two Majorana--Weyl spinors as follows:

\begin{equation}
  \epsilon
  =
  \epsilon^{1}+\epsilon^{2}\,,
  \hspace{1cm}
  \Gamma_{11}\epsilon^{i} =
  \left\{
    \begin{array}{lr}
      \sigma^{3\, i}{}_{j}\epsilon^{j}\,, \,\,\,\, & (2A)\,, \\
      \epsilon^{i}\,, \,\,\,\, & (2B)\,,\\
    \end{array}
  \right.
\end{equation}

\noindent
and the same will be done with the gravitinos and the dilatinos, although the
chiralities of the components of latter are the opposite to those of
$\epsilon$ and $\psi_{\mu}$.

The supersymmetry transformation rules of the chiral gravitinos and dilatinos
(indices now shown) are, with these conventions\footnote{Observe that all the
  $\Gamma_{11}$s in the transformations of Ref.~\cite{Bergshoeff:2001pv} have
  disappeared because we are using chiral spinors.}

\begin{subequations}
  \begin{align}
    \delta_{\epsilon}\psi_{\mu}
    & =
      \nabla_{\mu}\epsilon -\tfrac{1}{8}\not\!\!H_{\mu}\sigma^{3}\epsilon
      +\tfrac{1}{16}e^{\phi}\sum_{n=0}^{10}\frac{i^{n}}{n!}\not\!\!G^{(n)}
      \Gamma_{\mu}\mathcal{P}_{(n)}\epsilon\,,
    \\
    & \nonumber \\
    \delta_{\epsilon}\lambda
    & =
      \left(\not\!\partial \phi -\tfrac{1}{12}\not\!\!H\sigma^{3}\right)\epsilon
      +\tfrac{1}{8}e^{\phi}\sum_{n=0}^{10}\frac{(-i)^{n}(5-n)}{n!}\not\!\!G^{(n)}
      \mathcal{P}_{(n)}\epsilon\,,
  \end{align}
\end{subequations}

\noindent
where

\begin{equation}
  \mathcal{P}_{(n)}
  = 
  \left\{
    \begin{array}{lr}
      i\sigma^{1}\,, \,\,\,\,\, n=0,3,4,7,8\,, \,\,\,\, &
       (\text{when}\,\,\,\, (-1)^{[(n+1)/2]}=+1)\\
      & \\
      \sigma^{2}\,, \,\,\,\,\, n=1,2,5,6,9,10\,,  \,\,\,\, &
       (\text{when}\,\,\,\, (-1)^{[(n+1)/2]}=-1)\\
    \end{array}
  \right.
\end{equation}

With a commuting spinor $\epsilon =(\epsilon^{i})$, $i=1,2$ we can construct
the following spinor bilinears:

\begin{itemize}

\item A set of 1-, 5- and 9-forms

\begin{equation}
  K^{(1,5,9)} \equiv i \epsilon^{i}\Gamma^{(1,5,9)}\epsilon^{i}\,.
\end{equation}

\item Another set of 1-, 5- and 9-forms

\begin{equation}
  \tilde{K}^{(1,5,9)} \equiv i \sigma^{3}_{ij}\epsilon^{i}\Gamma^{(1,5,9)}\epsilon^{j}\,.
\end{equation}

\item Eleven $n$-forms

\begin{equation}
  \Omega^{(n)} \equiv i^{n-1} \mathcal{P}_{(n+2)ij}\epsilon^{i}\Gamma^{(1)}\epsilon^{j}\,.
\end{equation}

\end{itemize}

Not all of them are independent. Some of them are related by Hodge duality:

\begin{equation}
  \label{eq:dualitiesbetweenbilinears}
    \begin{aligned}
      \mathbf{2A}:\,\,\, & 
      \star K^{(9)} = \tilde{K}^{(1)}\,,\,\,\,
      \star \tilde{K}^{(9)} = K^{(1)}\,,\,\,\,   
      \star K^{(5)} = \tilde{K}^{(5)}\,,\,\,\,
                           \star \tilde{K}^{(5)} = K^{(5)}\,,\,\,\,\\
      & \\
      \mathbf{2B}:\,\,\, & 
      \star K^{(9)} = K^{(1)}\,,\,\,\,
      \star \tilde{K}^{(9)} = \tilde{K}^{(1)}\,,\,\,\,   
      \star K^{(5)} = K^{(5)}\,,\,\,\,
                           \star \tilde{K}^{(5)} = \tilde{K}^{(5)}\,,\,\,\,   \\
      & \\
 \text{and}\,\,\, &\Omega^{(10-n)}= (-1)^{[n/2]}\star \Omega^{(n)}\,.  
    \end{aligned}
\end{equation}

When $\epsilon$ is a Killing spinor (KS) \textit{i.e.}~when
$\delta_{\epsilon}\psi_{\mu} = \delta_{\epsilon} \lambda = 0$ (the Killing
spinor equations (KSEs)) these bilinears satisfy, among other, the following
first-order differential equations that we write, following
Ref.~\cite{Tomasiello:2011eb} in terms of the polyforms of RR field strengths
defined above and in terms of the polyform of bilinears
$\Omega \equiv \sum_{n=0}^{10}\Omega^{(n)}$:
  
\begin{subequations}
  \begin{align}
    \nabla_{(\mu}K^{(1)}_{\nu)} & = 0\,, \\
                                & \nonumber \\
    \imath_{K^{(1)}}H -d\tilde{K}^{(1)} & = 0\,,  \\
                                & \nonumber \\    
    \imath_{K^{(1)}}G +\tilde{K}^{(1)}\wedge G +d_{H^{-}}(e^{-\phi}\Omega) & = 0\,.
  \end{align}
\end{subequations}

The first equation just says that $K^{(1)}$ is a Killing vector. With that
information, the other equations can be easily recognized as the KR and RR
momentum map equations associated to to $K^{(1)}$
Eqs.~(\ref{eq:KRmomentummapeq}) and (\ref{eq:RRgeneralmomentummapeq}) with the
precise identifications

\begin{equation}
  \label{eq:identificationofmomentummaps}
  \tilde{K}^{(1)} = -P_{B\, K^{(1)}}^{(1)}\,,\hspace{1cm}
  e^{-\phi}\Omega^{(n)} =  P_{K^{(1)}}^{(n)}\,.
\end{equation}

The generalized Komar charge associated to the Killing vector $K^{(1)}$
involves the exterior derivative of the 1-form dual to it $K^{(1)}$,  which
was not given in Ref.~\cite{Tomasiello:2011eb}. Using the KSEs it can be shown
that

\begin{equation}
    dK^{(1)} - \imath_{\tilde{K}^{(1)}}H +\tfrac{1}{4} e^{\phi}
    \sum_{n=0}^{8} (-1)^{n} \left[\imath_{G^{(n)}}\Omega^{(n+2)}
      +\imath_{\Omega^{(n)}} G^{(n+2)}
   \right] =0\,. 
\end{equation}

Now, taking the Hodge dual of the whole expression and using the general
formula for $p$ and $q$-forms\footnote{By definition, for any $p$- and $q$-forms with $p>q$
  \begin{equation}
    \imath_{A^{(q)}} B^{(p)}
    \equiv
    \frac{1}{q! (p-q)!} A^{(q)\, \mu_{1}\cdots \mu_{q} }
    B^{(p)}_{\mu_{1}\cdots \mu_{q}\alpha_{1}\cdots \alpha_{p-q}}
    dx^{\alpha_{1}}\wedge \cdots \wedge  dx^{\alpha_{p-q}}\,.
  \end{equation}
}

\begin{equation}
  \begin{aligned}
    \star \imath_{A^{(q)}} B^{(p)}
    & = (-1)^{q(d-p)} A^{(q)}\wedge \star B^{(p)}\,, \\
  \end{aligned}  
\end{equation}

\noindent
we get
  
\begin{equation}
  \star dK^{(1)} +\tilde{K}^{(1)}\wedge \star H
  +\tfrac{1}{4} e^{\phi} \sum_{n=0}^{8}\left[G^{(n)} \wedge \star \Omega^{(n+2)}
      +\Omega^{(n)} \wedge \star G^{(n+2)}
   \right] = 0\,. 
\end{equation}

Multiplying the whole expression by $e^{-2\phi}$ and using now the duality
constraints Eqs.~(\ref{eq:dualitiesbetweenbilinears}) and the identification
of momentum maps Eq.~(\ref{eq:identificationofmomentummaps}) we find that the
above expression is just the generalized Komar charge associated to the
supersymmetric Killing vector, which implies that

\begin{equation}
  \mathbf{K}[K^{(1)}]
  =
  0\,,
  \,\,\,\,\,
  \text{Q.E.D.}
\end{equation}

\section{Conclusions}
\label{sec-discussion}

In this work we have computed the generalized Komar charges of the
$\mathcal{N}=2A,B, d=10$ supergravity theories. They are on-shell
closed, manifestly gauge-invariant 8-forms whose structure is that of
a linear combination of exterior products of field strengths with
momentum maps, which are gauge-invariant by construction.

While this expression is manifestly gauge invariant, it is not always
the most convenient because the field strengths are not closed on
shell and cannot always be interpreted as charges. On the other hand,
the momentum maps are not closed on Killing horizons, a property that
leads to restricted forms of the generalized zeroth law of black hole
mechanics \cite{Elgood:2020svt,Elgood:2020mdx,Elgood:2020nls}.  

We have shown, though, that these generalized Komar charges can be
rewritten in an entirely equivalent form as a linear combination of
exterior products of Page charges (on-shell closed but not manifestly
gauge invariant) and combinations of momentum maps which are closed on
Killing horizons and can be understood as potentials. This form should
immediately allow us to derive Smarr formulas, although, when
higher-rank forms are involved, there are other technical problems
that should be addressed first \cite{kn:BCOR}.

Finally, we have shown that, for supersymmetric solutions, the
generalized Komar charges of these theories associated to the
supersymmetric Killing vector vanish identically if we choose as
momentum maps certain spinor bilinears. Momentum maps are defined up
to closed forms and, since the equations that define the higher-rank
ones involve the lower-rank ones, this ambiguity can modify those
equations in a highly non-trivial way \cite{Gibbons:2013tqa}. It is a
highly non-trivial fact that those spinor bilinears satisfy, precisely,
the unmodified momentum-map equations.

These results suggest several important questions on which we are
currently working:

\begin{enumerate}
\item Why are the spinor bilinears associated to momentum maps? Is
  this a general fact? \cite{kn:MO}

\item Why does the generalized Komar charge vanish for supersymmetric
  solutions? Is this a general property of all supergravity
  theories. \cite{kn:MO}\footnote{A recent paper that addresses the
    relation between the equivariant localization formalism and
    supersymmetry is \cite{Galli:2026lsl}.}

\item How can we exploit these identities in brane solutions?
  \cite{kn:BCOR} (Generalized Komar integrals typically diverge for
  brane solutions.)

\item How can we exploit these results in the evaluation of the
  Euclidean action of supersymmetric and non-supersymmetric solutions?
  \cite{kn:CO} 
  
\end{enumerate}

Work in these directions is well in progress and will soon be reported.

\section*{Disclaimer}

No piece of this work has been carried out by or with the help of AI systems.

\section*{Acknowledgments}

We would like to thank useful and enlightening conversations with I.~Bandos,
J.L.V.~Cerdeira, P.~Meessen and J.~Sparks and the CERN TH group for its
financial support and hospitality (TO) in the last stages of this work. The
work of GB and TO has been supported in part by the MCI, AEI, FEDER (UE) grant
PID2024-155685NB-C21 (``Gravity, Supergravity and Superstrings'' (GRASS)). The
work of GB has been supported by the fellowship CEX2020-001007-S-20-5. This
publication has been funded within the framework of the R\&D\&I Project
CEX2025-001574-S, funded by MICIU/AEI/10.13039/501100011033. The research
presented in this publication falls within the research line \textit{Strings
  and Quantum Gravity}. TO wishes to thank M.M.~Fern\'andez for her permanent
support.

\appendix



\begin{thebibliography}{99}

\bibitem{Regge:1974zd}
T.~Regge and C.~Teitelboim,
``Role of Surface Integrals in the Hamiltonian
Formulation of General Relativity,''
Annals Phys. \textbf{88} (1974), 286
\doi{10.1016/0003-4916(74)90404-7}

\bibitem{Banks:1988yz}
T.~Banks and L.~J.~Dixon,
``Constraints on String Vacua with Space-Time Supersymmetry,''
Nucl. Phys. B \textbf{307} (1988), 93-108
\doi{10.1016/0550-3213(88)90523-8}

\bibitem{Banks:2010zn}
T.~Banks and N.~Seiberg,
``Symmetries and Strings in Field Theory and Gravity,''
Phys. Rev. D \textbf{83} (2011), 084019
\doi{10.1103/PhysRevD.83.084019}
[\arxiv{1011.5120} [hep-th]].

\bibitem{Abbott:1981ff}
L.~F.~Abbott and S.~Deser,
``Stability of Gravity with a Cosmological Constant,''
Nucl. Phys. B \textbf{195} (1982), 76-96
\doi{10.1016/0550-3213(82)90049-9}

\bibitem{Abbott:1982jh}
L.~F.~Abbott and S.~Deser,
``Charge Definition in Nonabelian Gauge Theories,''
Phys. Lett. B \textbf{116} (1982), 259-263
\doi{10.1016/0370-2693(82)90338-0}

\bibitem{Barnich:2001jy}
G.~Barnich and F.~Brandt,
``Covariant theory of asymptotic symmetries,
conservation laws and central charges,''
Nucl. Phys. B \textbf{633} (2002), 3-82
\doi{10.1016/S0550-3213(02)00251-1}
[\hepth{0111246} [hep-th]].

\bibitem{Barnich:2003xg}
G.~Barnich,
``Boundary charges in gauge theories: Using Stokes theorem in the bulk,''
Class. Quant. Grav. \textbf{20} (2003), 3685-3698
\doi{10.1088/0264-9381/20/16/310}
[\hepth{0301039} [hep-th]].

\bibitem{Komar:1958wp}
A.~Komar,
``Covariant conservation laws in general relativity,''
Phys. Rev. \textbf{113} (1959), 934-936
\doi{10.1103/PhysRev.113.934}

\bibitem{Bardeen:1973gs}
J.~M.~Bardeen, B.~Carter and S.~W.~Hawking,
``The Four laws of black hole mechanics,''
Commun. Math. Phys. \textbf{31} (1973), 161-170
\doi{10.1007/BF01645742}

\bibitem{Carter:1973rla}
B.~Carter,
``Black holes equilibrium states,''
Contribution to: Les Houches Summer School of Theoretical Physics, 57-214.

\bibitem{Magnon:1985sc}
A.~Magnon,
``On Komar integrals in asymptotically anti-de Sitter space-times,''
J. Math. Phys. \textbf{26} (1985), 3112-3117
\doi{10.1063/1.526690}

\bibitem{Bazanski:1990qd}
S.~L.~Bazanski and P.~Zyla,
``A Gauss type law for gravity with a cosmological constant,''
Gen. Rel. Grav. \textbf{22} (1990), 379-387

\bibitem{Kastor:2008xb}
D.~Kastor,
``Komar Integrals in Higher (and Lower) Derivative Gravity,''
Class. Quant. Grav. \textbf{25} (2008), 175007
\doi{10.1088/0264-9381/25/17/175007}
[\arxiv{0804.1832} [hep-th]].

\bibitem{Smarr:1972kt}
L.~Smarr,
``Mass formula for Kerr black holes,''
Phys. Rev. Lett. \textbf{30} (1973), 71-73
[erratum: Phys. Rev. Lett. \textbf{30} (1973), 521-521]
\doi{10.1103/PhysRevLett.30.71}

\bibitem{Ballesteros:2023iqb}
R.~Ballesteros, C.~G\'omez-Fayr\'en, T.~Ort\'{\i}n and M.~Zatti,
``On scalar charges and black hole thermodynamics,''
JHEP \textbf{05} (2023), 158
\doi{10.1007/JHEP05(2023)158}
[\arxiv{2302.11630} [hep-th]].

\bibitem{Ballesteros:2024prz}
R.~Ballesteros and T.~Ort\'{\i}n,
``Generalized Komar charges and Smarr formulas
for black holes and boson stars,''
SciPost Phys. Core \textbf{8} (2025) no.038, 038
\doi{10.21468/SciPostPhysCore.8.2.038}
[\arxiv{2409.08268} [gr-qc]].

\bibitem{Barbagallo:2026hof}
G.~Barbagallo and T.~Ort\'{\i}n,
``Derivation of the Smarr formula from the Komar charge
in Einstein-nonlinear electrodynamics theories and
applications to regular black holes,''
[\arxiv{2605.02813} [gr-qc]].

\bibitem{Liberati:2015xcp}
S.~Liberati and C.~Pacilio,
``Smarr Formula for Lovelock Black Holes: a Lagrangian approach,''
Phys. Rev. D \textbf{93} (2016) no.8, 084044
\doi{10.1103/PhysRevD.93.084044}
[\arxiv{1511.05446} [gr-qc]].

\bibitem{Ortin:2021ade}
T.~Ort\'{\i}n,
``Komar integrals for theories of higher order
in the Riemann curvature and black-hole chemistry,''
JHEP \textbf{08} (2021), 023
\doi{10.1007/JHEP08(2021)023}
[\arxiv{2104.10717} [gr-qc]].

\bibitem{Ortin:2022uxa}
T.~Ort\'{\i}n and D.~Pere\~niguez,
``Magnetic charges and Wald entropy,''
JHEP \textbf{11} (2022), 081
\doi{10.1007/JHEP11(2022)081}
[\arxiv{2207.12008} [hep-th]].

\bibitem{Cerdeira:2025elp}
J.~L.~V.~Cerdeira and T.~Ort\'{\i}n,
``On-shell Lagrangians as total derivatives
and the generalized Komar charge,''
JHEP \textbf{09} (2025), 068
\doi{10.1007/JHEP09(2025)068}
[\arxiv{2506.14024} [hep-th]].

\bibitem{Mitsios:2021zrn}
D.~Mitsios, T.~Ort\'{\i}n and D.~Pere\~n\'{\i}guez,
``Komar integral and Smarr formula for axion-dilaton
black holes versus S duality,''
JHEP \textbf{08} (2021), 019
\doi{10.1007/JHEP08(2021)019}
[\arxiv{2106.07495} [hep-th]].

\bibitem{Meessen:2022hcg}
P.~Meessen, D.~Mitsios and T.~Ort\'{\i}n,
``Black hole chemistry, the cosmological constant
and the embedding tensor,''
JHEP \textbf{12} (2022), 155
\doi{10.1007/JHEP12(2022)155}
[\arxiv{2203.13588} [hep-th]].

\bibitem{Ballesteros:2023muf}
R.~Ballesteros and T.~Ort\'{\i}n,
``Hairy black holes, scalar charges and extended thermodynamics,''
Class. Quant. Grav. \textbf{41} (2024) no.5, 055007
\doi{10.1088/1361-6382/ad210a}
[\arxiv{2308.04994} [gr-qc]].

\bibitem{Barbagallo:2025qdy}
  G.~Barbagallo, J.~L.~V.~Cerdeira, C.~G\'omez-Fayr\'en,
  P.~Meessen and T.~Ort\'{\i}n,
  ``Higher-form symmetries in supergravity, scalar charges
  and black-hole thermodynamics,''
JHEP \textbf{06} (2026), 235
\doi{10.1007/JHEP06(2026)235}
[\arxiv{2512.22565} [hep-th]].

\bibitem{Bandos:2023zbs}
I.~Bandos and T.~Ort\'{\i}n,
``Noether-Wald charge in supergravity: the fermionic contribution,''
JHEP \textbf{12} (2023), 095
\doi{10.1007/JHEP12(2023)095}
[\arxiv{2305.10617} [hep-th]].

\bibitem{Bandos:2024rit}
I.~Bandos, P.~Meessen and T.~Ort\'{\i}n,
``Komar charge of $ \mathcal{N} $ = 2 supergravity
and its superspace generalization,''
JHEP \textbf{05} (2025), 231
\doi{10.1007/JHEP05(2025)231}
[\arxiv{2412.18510} [hep-th]].

\bibitem{BenettiGenolini:2023kxp}
P.~Benetti Genolini, J.~P.~Gauntlett and J.~Sparks,
``Equivariant Localization in Supergravity,''
Phys. Rev. Lett. \textbf{131} (2023) no.12, 121602
doi:10.1103/PhysRevLett.131.121602
[arXiv:2306.03868 [hep-th]].

\bibitem{Couzens:2026xmi}
C.~Couzens, A.~L{\"u}scher and J.~Sparks,
``Ten-dimensional localization,''
[\arxiv{2607.05529} [hep-th]].

\bibitem{BenettiGenolini:2026kjy}
P.~Benetti Genolini, F.~Gaar, J.~P.~Gauntlett, J.~Park and J.~Sparks,
``Localizing the M-Theory Path Integral,''
[\arxiv{2609.31829} [hep-th]].

\bibitem{Ortin:2026fzm}
T.~Ort\'{\i}n,
``On the generalized Komar charge of supersymmetric solutions,''
[\arxiv{2605.05178} [hep-th]].

\bibitem{Tomasiello:2011eb}
A.~Tomasiello,
``Generalized structures of ten-dimensional supersymmetric solutions,''
JHEP \textbf{03} (2012), 073
\doi{10.1007/JHEP03(2012)073}
\arxiv{1109.2603} [hep-th].

\bibitem{Bergshoeff:2001pv}
E.~Bergshoeff, R.~Kallosh, T.~Ort\'{\i}n, D.~Roest and A.~Van Proeyen,
``New formulations of D = 10 supersymmetry and D8 - O8 domain walls,''
Class. Quant. Grav. \textbf{18} (2001), 3359-3382
\doi{10.1088/0264-9381/18/17/303}
\hepth{0103233} [hep-th]].

\bibitem{Ortin:2015hya}
T.~Ort\'{\i}n,
``Gravity and Strings,''
Cambridge University Press, 2015,
ISBN 978-0-521-76813-9, 978-0-521-76813-9, 978-1-316-23579-9
\doi{10.1017/CBO9781139019750}
  
\bibitem{Giani:1984wc}
F.~Giani and M.~Pernici,
``N=2 Supergravity in ten-dimensions,''
Phys. Rev. D \textbf{30} (1984), 325-333
\doi{10.1103/PhysRevD.30.325}
  
\bibitem{Campbell:1984zc}
I.~C.~G.~Campbell and P.~C.~West,
``N=2 D=10 Nonchiral Supergravity and Its Spontaneous Compactification,''
Nucl. Phys. B \textbf{243} (1984), 112-124
\doi{10.1016/0550-3213(84)90388-2}
\bibitem{Lee:1990nz}
J.~Lee and R.~M.~Wald,
``Local symmetries and constraints,''
J. Math. Phys. \textbf{31} (1990), 725-743
\doi{10.1063/1.528801}

\bibitem{Wald:1993nt}
R.~M.~Wald,
``Black hole entropy is the Noether charge,''
Phys. Rev. D \textbf{48} (1993) no.8, R3427-R3431
\doi{10.1103/PhysRevD.48.R3427}
[\grqc{9307038} [gr-qc]].

\bibitem{Iyer:1994ys}
V.~Iyer and R.~M.~Wald,
``Some properties of Noether charge and a
proposal for dynamical black hole entropy,''
Phys. Rev. D \textbf{50} (1994), 846-864
\doi{10.1103/PhysRevD.50.846}
[\grqc{9403028} [gr-qc]].

\bibitem{Elgood:2020svt}
Z.~Elgood, P.~Meessen and T.~Ort\'{\i}n,
``The first law of black hole mechanics in
the Einstein-Maxwell theory revisited,''
JHEP \textbf{09} (2020), 026
\doi{10.1007/JHEP09(2020)026}
[\arxiv{2006.02792} [hep-th]].

\bibitem{Elgood:2020mdx}
Z.~Elgood, D.~Mitsios, T.~Ort\'{\i}n and D.~Pere\~niguez,
``The first law of heterotic stringy black hole mechanics
at zeroth order in \ensuremath{\alpha}',''
JHEP \textbf{07} (2021), 007
\doi{10.1007/JHEP07(2021)007}
[\arxiv{2012.13323} [hep-th]].

\bibitem{Elgood:2020nls}
Z.~Elgood, T.~Ort\'{\i}n and D.~Pere\~niguez,
``The first law and Wald entropy formula of heterotic
stringy black holes at first order in $\alpha'$,''
JHEP \textbf{05} (2021), 110
\doi{10.1007/JHEP05(2021)110}
[\arxiv{2012.14892} [hep-th]].

\bibitem{kn:Lich}
A.~Lichnerowicz,
``Spineurs harmoniques,''
C.\ R.\ Acad.\ Sci.\ Paris {\bf 257} (1963) 7-9.

\bibitem{kn:Kos}
Y.~Kosmann,
``D\'eriv\'ees de Lie des spineurs,''
C.\ R.\ Acad.\ Sci.\ Paris S\'er. A {\bf 262} (1966) A289-A292.
                
\bibitem{kn:Kos2}
  Y.~Kosmann,
``D\'eriv\'ees de Lie des spineurs,''
Annali Mat.\ Pura Appl.\ {\bf (IV) 91} (1972) 317-395.

\bibitem{Hurley:cf}
D.~J.~Hurley and M.~A.~Vandyck,
``On The Concepts Of Lie And Covariant Derivatives Of Spinors. Part 1,''
J.\ Phys.\ A {\bf 27} (1994) 4569.

\bibitem{Figueroa-O'Farrill:1999va}
J.~M.~Figueroa-O'Farrill,
``On the supersymmetries of anti de Sitter vacua,''
Class.\ Quant.\ Grav.\  {\bf 16} (1999) 2043
[\hepth{9902066}].

\bibitem{Ortin:2002qb}
T.~Ort\'{\i}n,
``A Note on Lie-Lorentz derivatives,''
Class.\ Quant.\ Grav.\  {\bf 19} (2002) L143
\doi{10.1088/0264-9381/19/15/101}
[\hepth{0206159}].

\bibitem{Godina:2003tc}
M.~Godina and P.~Matteucci,
``Reductive G structures and Lie derivatives,''
J. Geom. Phys. \textbf{47} (2003), 66-86
\doi{10.1016/S0393-0440(02)00174-2}
[\matharxiv{0201235} [math.DG]].

\bibitem{Fatibene2011}
L.~Fatibene and M.~Francaviglia,
``General theory of lie derivatives for Lorentz tensors,''
Communications in Mathematics, {\bf19} (2011)  11.
[\arxiv{0904.0258} [math-ph]]

\bibitem{Prabhu:2015vua}
K.~Prabhu,
``The First Law of Black Hole Mechanics for Fields
with Internal Gauge Freedom,''
Class. Quant. Grav. \textbf{34} (2017) no.3, 035011
\doi{10.1088/1361-6382/aa536b}
[\arxiv{1511.00388} [gr-qc]].

\bibitem{Bunk:2023ojb}
S.~Bunk, V.~Mu\~noz and C.~S.~Shahbazi,
``The local moduli space of the Einstein-Yang-Mills system,''
[\arxiv{2311.07572} [math.DG]].

\bibitem{Jacobson:2015uqa}
T.~Jacobson and A.~Mohd,
``Black hole entropy and Lorentz-diffeomorphism Noether charge,''
Phys. Rev. D \textbf{92} (2015), 124010
\doi{10.1103/PhysRevD.92.124010}
[\arxiv{1507.01054} [gr-qc]].

\bibitem{Gibbons:2013tqa}
G.~W.~Gibbons and N.~P.~Warner,
``Global structure of five-dimensional fuzzballs,''
Class. Quant. Grav. \textbf{31} (2014), 025016
\doi{10.1088/0264-9381/31/2/025016}
[\arxiv{1305.0957} [hep-th]].

\bibitem{Romans:1985tz}
L.~J.~Romans,
``Massive N=2a Supergravity in Ten-Dimensions,''
Phys. Lett. B \textbf{169} (1986), 374
\doi{10.1016/0370-2693(86)90375-8}

\bibitem{Bergshoeff:1996ui}
E.~Bergshoeff, M.~de Roo, M.~B.~Green, G.~Papadopoulos and P.~K.~Townsend,
``Duality of type II 7 branes and 8 branes,''
Nucl. Phys. B \textbf{470} (1996), 113-135
\doi{10.1016/0550-3213(96)00171-X}
[\hepth{9601150} [hep-th]].

\bibitem{Aurilia:1980xj}
A.~Aurilia, H.~Nicolai and P.~K.~Townsend,
``Hidden Constants: The Theta Parameter of QCD and
the Cosmological Constant of N=8 Supergravity,''
Nucl. Phys. B \textbf{176} (1980), 509-522
\doi{10.1016/0550-3213(80)90466-6}

\bibitem{Duff:1980qv}
M.~J.~Duff and P.~van Nieuwenhuizen,
``Quantum Inequivalence of Different Field Representations,''
Phys. Lett. B \textbf{94} (1980), 179-182
\doi{10.1016/0370-2693(80)90852-7}

\bibitem{Zatti:2024vbz}
M.~Zatti,
``Exploring String Theory Solutions: Black  hole thermodynamics
with $\alpha'$ corrections and type II  compactifications,''
PhD Thesis, U.~Aut\'onoma de Madrid, 2024.
[\arxiv{2505.01191} [hep-th]].

\bibitem{Schwarz:1983qr}
J.~H.~Schwarz,
``Covariant Field Equations of Chiral N=2 D=10 Supergravity,''
Nucl. Phys. B \textbf{226} (1983), 269
\doi{10.1016/0550-3213(83)90192-X}
  
\bibitem{Howe:1983sra}
P.~S.~Howe and P.~C.~West,
``The Complete N=2, D=10 Supergravity,''
Nucl. Phys. B \textbf{238} (1984), 181-220
\doi{10.1016/0550-3213(84)90472-3}

\bibitem{Pasti:1996vs}
P.~Pasti, D.~P.~Sorokin and M.~Tonin,
``On Lorentz invariant actions for chiral p forms,''
Phys. Rev. D \textbf{55} (1997), 6292-6298
\doi{10.1103/PhysRevD.55.6292}
[\hepth{9611100} [hep-th]].

\bibitem{Sen:2015nph}
A.~Sen,
``Covariant Action for Type IIB Supergravity,''
JHEP \textbf{07} (2016), 017
\doi{10.1007/JHEP07(2016)017}
[\arxiv{1511.08220} [hep-th]].

\bibitem{Bergshoeff:1995sq}
E.~Bergshoeff, H.~J.~Boonstra and T.~Ort\'{\i}n,
``S duality and dyonic p-brane solutions in type II string theory,''
Phys. Rev. D \textbf{53} (1996), 7206-7212
\doi{10.1103/PhysRevD.53.7206}
[\hepth{9508091} [hep-th]].

\bibitem{Fernandez-Melgarejo:2023kwk}
  J.~J.~Fern\'andez-Melgarejo, G.~Giorgi, C.~G\'omez-Fayr\'en,
  T.~Ort\'{\i}n and M.~Zatti,
``Democratic actions with scalar fields: Symmetric sigma models,
supergravity actions and the effective theory of the type IIB superstring,''
SciPost Phys. Core \textbf{7} (2024) no.4, 068
\doi{10.21468/SciPostPhysCore.7.4.068}
[\arxiv{2401.00549} [hep-th]].

\bibitem{kn:BCOR}
  G.~Barbagallo, J.L.V.~Cerdeira, T.~Ort\'{\i}n and A.~Ruip\'erez,
  work in progress.

\bibitem{kn:MO}
P.~Meessen and T.~Ort\'{\i}n,
work in progress  

\bibitem{Galli:2026lsl}
M.~Galli, C.~Kennedy, P.~Raina and G.~Tartaglino-Mazzucchelli,
``Superform Approach to Equivariant Localization in Supergravity,''
[\arxiv{2605.17919} [hep-th]].

\bibitem{kn:CO}
J.L.V.~Cerdeira and T.~Ort\'{\i}n, to appear.

\end{thebibliography}
\end{document}